\documentclass[a4paper,11pt]{article}
\pdfoutput=1 

\usepackage{jheppub} 
\usepackage{bbm,bm,graphicx,mathtools,color,slashed,hyperref}
\usepackage{amssymb,amsmath}
\usepackage{braket}
\usepackage{subfig}

\newcommand{\diff}{\mathrm{d}}

\newcommand{\Diff}{{\mathcal{D}}}

\newcommand{\tr}{\mathrm{tr}}
\newcommand{\im}{\mathrm{i}}

\newcommand{\rme}{\mathrm{e}}

\title{
Wilson-'t Hooft classification and the perimeter law for dyonic loops in 3d monopole semiclassics
}

\author{Yui Hayashi and}
\emailAdd{yui.hayashi@yukawa.kyoto-u.ac.jp}
\affiliation{Yukawa Institute for Theoretical Physics, Kyoto University, Kyoto 606-8502, Japan}

\author{Yuya Tanizaki}
\emailAdd{yuya.tanizaki@yukawa.kyoto-u.ac.jp}

\preprint{YITP-25-197}

\abstract{
We investigate the long-distance behavior of dyonic loop operators in 4d $SU(N)$ gauge theories on $\mathbb{R}^3 \times S^1$ using the 3d monopole semiclassics. 
If we employ the naive definition of the 't Hooft loop in the Abelianized regime, the dyonic loop operators do not admit the well-defined computations within the effective field theory. 
Moreover, if one forcibly proceeds with the computations of their expectation values, all the dyonic loops turn out to show the area law, which contradicts the prediction of the Wilson-'t Hooft classification. 
In this paper, we resolve this puzzle by employing the notion of screening for line operators, and we argue that the dyonic loops are screened by a defect known as the twist vortex, which is non-dynamical in the infrared effective theory but is dynamical in the original ultraviolet theory. 
The dyonic loops properly dressed by twist vortices admit the well-defined computations within the effective field theory, and we reproduce the kinematic prediction of the Wilson-'t~Hooft classification using the $3$d monopole semiclassics.
Furthermore, we apply our framework to the thermal deconfined phase to evaluate the dual string tension, elucidating the topological nature of $\mathbb{Z}_N$ domain walls. 
We confirm that the domain-wall state has the phase transition at $\theta=\pi$ in the thermal deconfined phase despite the fact that the bulk state is smooth there. 
}

\begin{document}

\maketitle

\section{Introduction}

Color confinement is an important and fundamental property of $4$d Yang-Mills theory, which explains why one can only observe color-singlet hadrons in our universe~\cite{Wilson:1974sk}. 
One of the standard scenarios for the confinement mechanism is the dual Meissner effect~\cite{Nambu:1974zg, Mandelstam:1974pi, Polyakov:1975rs, tHooft:1977nqb}: 
While all the fundamental fields in the gauge-theory Lagrangian carry electric charges, the $4$d gauge theories also contain magnetically charged particles in their spectrum~\cite{tHooft:1974kcl, Polyakov:1974ek, Goddard:1976qe}. 
When the gauge couplings become sufficiently strong, those magnetic particles become light objects and start to condense in the vacuum, which causes the confinement of the electric flux.  

Once we accept this intuitive understanding for the confinement mechanism, we notice that there are a wide variety of the possible electric-magnetic condensations that cause a non-zero mass gap. 
This pursuit naturally motivates a broader objective, i.e., the classification of all possible gapped phases of $4$d $SU(N)$ gauge theories with adjoint matter: What is the gauge-invariant order parameter to distinguish them? 
A key framework for this task is the Wilson-'t Hooft classification \cite{Wilson:1974sk, tHooft:1977nqb, tHooft:1979rtg, tHooft:1981bkw}, which proposes to distinguish different phases by examining the long-distance behavior of dyonic probe particles. 
We identify their long-distance behaviors via the area law or the perimeter law for the $\mathbb{Z}_N\times \mathbb{Z}_N$ set of Wilson and 't Hooft line operators.

From the modern perspective of generalized global symmetries \cite{Aharony:2013hda, Kapustin:2014gua, Gaiotto:2014kfa}, the $4$d $SU(N)$ Yang-Mills theory with adjoint matter enjoys the $\mathbb{Z}_N$ $1$-form symmetry (denoted by $\mathbb{Z}_N^{[1]}$). 
Then, Wilson loops are genuine line operators charged under $\mathbb{Z}_N^{[1]}$, while
the 't Hooft loops are considered non-genuine line operators, which live on the boundary of the $\mathbb{Z}_N^{[1]}$ generator defined on an open codimension-$2$ surface. 
To diagnose if $\mathbb{Z}_N^{[1]}$ is spontaneously broken or not, we should check the perimeter or area law of Wilson loops, and thus the genuine line operators are enough. 
When some of the Wilson loops obey the area law, however, we obtain finer classification of the confinement phases by using non-genuine loop operators according to the Wilson-'t~Hooft classification. 
Nowadays, the utility of such non-genuine operators for phase classification is well-recognized, and one of the most famous examples would be string-order parameters to detect symmetry-protected topological (SPT) phases in $(1+1)$d~\cite{denNijs:1989ntw, Kennedy:1992tke, Kennedy:1992ifl, Li:2023ani}. 
This viewpoint, which bridges the traditional Wilson-'t Hooft picture with the modern framework of generalized symmetries, has been recently explored in Refs.~\cite{Nguyen:2023fun, Maeda:2025ycr}: Wilson-'t~Hooft classification characterizes both the unbroken subgroup of $\mathbb{Z}_N^{[1]}$ and the stacking of SPT phases of the unbroken symmetry. 
The primary focus of this paper is to investigate long-range behaviors of these dyonic lines, particularly on the non-genuine 't Hooft lines\footnote{Beyond the gapped phase classification, the 't~Hooft loops have been explored in various contexts.
The (non-genuine) 't~Hooft loop and its ``dual string tension'' in deconfined phase have been investigated in lattice studies \cite{deForcrand:2000fi, deForcrand:2004jt, deForcrand:2005pb, Bursa:2005yv}.
Also, in the deconfined phase, the ``dual string tension'' of the spatial 't~Hooft loop is almost equivalent to the $\mathbb{Z}_N$ domain wall tension \cite{Korthals-Altes:1999cqo}, so it has attracted interest in phenomenological contexts \cite{Bhattacharya:1992qb, Giovannangeli:2001bh, Giovannangeli:2002uv, Giovannangeli:2004sg, Hidaka:2009hs}. 
The 't~Hooft loops in $\mathcal{N}=4$ supersymmetric Yang-Mills theory are also studied in Refs.~\cite{Kapustin:2005py, Gomis:2009ir, Giombi:2009ek, Armoni:2008yp}.}, in the calculable regime of confining gauge theories.

In general, calculating observables is extremely hard in strongly coupled theories, and we would like to deform the theory into a weakly-coupled setup without encountering the phase transitions. 
One such deformation for 4d $SU(N)$ Yang-Mills theory is the $3$d monopole semiclassics~\cite{Unsal:2007jx, Unsal:2007vu, Unsal:2008ch} (see also \cite{Davies:1999uw, Davies:2000nw}), which deforms 4d gauge theories into weakly coupled confining theories on $\mathbb{R}^3 \times S^1$.
Upon $S^1$ compactification with center-stabilizing deformation (or in the presence of adjoint fermions with the periodic boundary condition), the 3d effective theory becomes the $U(1)^{N-1}$ gauge theory with monopoles and exhibits confinement by the Polyakov mechanism \cite{Polyakov:1975rs}.
Hence, at small $S^1$, the 4d gauge theory becomes the 3d weakly coupled confining theory, and it would provide an ideal setup for computing the dyonic line operators to explicitly confirm the Wilson-'t~Hooft classification.

The definition of 't~Hooft loops for the $3$d monopole semiclassics has been studied in Refs.~\cite{Anber:2015wha, Anber:2017rch} to understand the set of genuine line operators for the choice of the global structure of the gauge group. 
For our purpose of studying the gapped phases of the $SU(N)$ gauge theory, it would be natural to simply adapt this definition of the 't~Hooft loop as the non-genuine line operator and then we would be able to study the low-energy behaviors of dyonic line operators. 
If we try to do it, however, the dyonic lines in this conventional definition turn out to be ill-defined within the low-energy effective theory of the $3$d monopole semiclassics.
Furthermore, if one proceeds with the computation neglecting those singularities, all the dyonic lines turn out to show the area law in the $3$d monopole semiclassics, which contradicts the expectation from the Wilson-'t~Hooft classification.

In this paper, we resolve this puzzle by clarifying the importance of the screening of the dyonic loops by twist vortices. 
In general, there exists operator mixing between two different operators unless protected by symmetry, and we can employ any generic operators in the given symmetry class as an order parameter. 
However, quite often, such an operator mixing is accidentally prohibited within the low-energy effective theory due to an emergent symmetry, and the correct behavior cannot be obtained unless we use the properly dressed operators from the beginning. 
For example, while we know the $N$-th power of the Wilson loop $W^N(C)$ decays as the perimeter law in the $4$d $SU(N)$ gauge theories, its naive low-energy counterpart in the Abelianized effective theory shows the area law.
This discrepancy can be understood from the screening for the Wilson loop by heavy $W$-bosons, which are integrated out to obtain the Abelianized theory, and we should consider the dressed charge-$N$ Wilson loop from the beginning to reproduce the perimeter law.
We argue that the situation for the 't Hooft loop is basically identical, and we introduce the 't Hooft loop screened by twist vortices. 
We show that the screened 't~Hooft loop is the well-defined operator within the Abelianized effective theory, and, moreover, the screened 't Hooft loop gives the area-law or perimeter-law behavior consistent with the Wilson-'t Hooft classification.

This paper is organized as follows.
In Section~\ref{sec:review_and_problem}, we give a review on the Wilson-'t Hooft classification and on the 3d monopole semiclassics. We also mention the naive definition of the non-genuine spatial 't Hooft loop in the abelianized effective theory, highlighting the inconsistency between the naive definition and the prediction of the Wilson-'t Hooft classification.
In Section~\ref{sec:Screening_tHooftloop}, we propose the screening of the 't Hooft loop by the twist vortex. We show that the screened operator is well-defined in the infrared effective theory (only in terms of the dual photon).
In Section~\ref{sec:behavior_dyonicloops}, we discuss the behavior of dyonic lines in the confining phase. We confirm that the screened 't Hooft loop obeys the perimeter law, consistent with the Wilson-'t Hooft classification.
In Section~\ref{sec:tHooft_thermal_deconf}, we investigate the spatial 't Hooft loop in the thermal deconfined phase. We discuss the refined classification of the $\mathbb{Z}_N$ domain walls, which corresponds to that of the dyonic loops.
As an example, we apply the semiclassical description of the softly-broken $\mathcal{N}=1$ supersymmetric Yang-Mills (SYM) theory to mimic the thermal deconfined phase \cite{Poppitz:2012sw, Anber:2014lba, Chen:2020syd} and discuss the relationship between the domain-wall tension and the dual string tension of the spatial dyonic loops, aligning with the kinematical prediction. 
Section~\ref{sec:Summary} is devoted to summary and discussion.
In Appendix~\ref{app:Lattice}, we provide a lattice illustration of our proposal in a simplified setup of the $SU(N)$ lattice gauge theory which reduces to $U(1)^{N-1} \rtimes S_N$ at low energies. Using the standard definition of the 't Hooft loop (as the boundary of the 1-form symmetry generator employed in, e.g., \cite{deForcrand:2000fi}), we demonstrate that the screening by twist vortices is essential for the perimeter law. 


\section{Wilson-'t Hooft classification versus 3d monopole semiclassics}
\label{sec:review_and_problem}

In this section, we first give a review on the Wilson-'t~Hooft classification from the viewpoint of the $\mathbb{Z}_N^{[1]}$ symmetry to understand the general behaviors of the $4$d gapped quantum phases. 
Next, we review the 3d monopole semiclassics, which provides the weakly coupled description for the confinement on $\mathbb{R}^3 \times S^1$.
We then discuss the naive definition of the 't~Hooft loop operator in this framework and pose a puzzle: All the dyonic operators show confinement, which contradicts the expectation from the Wilson-'t Hooft classification.

\subsection{Review of the Wilson-'t Hooft classification and the \texorpdfstring{$\mathbb{Z}_N^{[1]}$}{ZN 1-form} symmetry}

$4$d $SU(N)$ Yang-Mills theories coupled with adjoint matters have the $\mathbb{Z}_N^{[1]}$ symmetry as the global symmetry~\cite{Gaiotto:2014kfa}. 
We can introduce the $\mathbb{Z}_N$ two-form background gauge field $B_{4\diff}$, which is realized here as the $U(1)$ two-form gauge field with the constraint $\int_{M_2} B_{4\diff}\in \frac{2\pi}{N}\mathbb{Z}$ for all the closed $2$-surfaces $M_2$~\cite{Kapustin:2014gua}. 
When introducing the background gauge field, there exists a local counterterm, $\im\frac{N k_{\mathrm{UV}}}{4\pi}\int B_{4\diff}\wedge B_{4\diff}$, with the discrete label $k_{\mathrm{UV}}\in \mathbb{Z}_N$, and we choose the ultraviolet (UV) regularization that is consistent with $k_{\mathrm{UV}}=0$ throughout this paper: 
The standard Wilson lattice regularization is such a regulator in the minimal coupling procedure. 

The $4$d gapped phases with the $\mathbb{Z}_N^{[1]}$ symmetry can be classified by two ingredients; 
\begin{itemize}
    \item Spontaneous symmetry breaking (SSB), $\mathbb{Z}_{N}^{[1]}\xrightarrow{\mathrm{SSB}}\mathbb{Z}_{n}^{[1]}$, with some divisor $n$ of $N$. 
    \item Stacking of the symmetry-protected topological (SPT) phase, $\frac{\im nk}{4\pi}\int B_{4\diff}\wedge B_{4\diff}$, for the unbroken $\mathbb{Z}_n^{[1]}$ symmetry. Here, the SPT level is characterized by $k\sim k+n$. 
\end{itemize}
The low-energy behavior of the partition function $Z[B_{4\diff}]$ with the flat background gauge field $B_{4\diff}$ can be  written as~\cite{Nguyen:2023fun, Maeda:2025ycr} (see also \cite{tHooft:1979rtg})
\begin{equation}
    Z[B_{4\diff}]=\int \Diff b_{N/n} \Diff a_{N/n} \exp\left(\frac{\im (N/n)}{2\pi}\int b_{N/n}\wedge (\diff a_{N/n} - n B_{4\diff})+\frac{\im n k}{4\pi}\int B_{4\diff}\wedge B_{4\diff}\right),
    \label{eq:Z_TQFT}
\end{equation}
up to the gravitational counterterm, where $a_{N/n},b_{N/n}$ are $U(1)$ $1$- and $2$-form gauge fields describing the $\mathbb{Z}_{N/n}$ topological gauge theory. 
Under the background $1$-form gauge transformation, $B_{4\diff}\mapsto B_{4\diff}+\diff \Lambda^{(1)}$, the dynamical fields $a_{N/n}$ and $b_{N/n}$ should transform as 
\begin{align}
    b_{N/n}\mapsto b_{N/n}+\frac{k}{N/n}\diff \Lambda^{(1)},\quad 
    a_{N/n}\mapsto a_{N/n}+n \Lambda^{(1)}. 
    \label{eq:TQFT_GaugeTransformation}
\end{align}
To identify the low-energy realization of the $\mathbb{Z}_N^{[1]}$ symmetry, we need to specify these labels $n$ and $k$ for the gapped quantum phases. 
The Wilson-'t~Hooft classification~\cite{Wilson:1974sk, tHooft:1977nqb} indicates that this is possible by identifying the dyonic lines showing the perimeter law~\cite{Nguyen:2023fun, Maeda:2025ycr}.

In the $SU(N)$ gauge theories with adjoint matters, there are two types of important loop operators, Wilson and 't~Hooft loops. 
The Wilson loops are the genuine line operators charged under the $\mathbb{Z}_N^{[1]}$ symmetry, and we write the fundamental Wilson loop as $W(C)$:
\begin{equation}
    W(C):=\frac{1}{N}\tr \left[\mathcal{P}\exp\left(\im \int_C a\right)\right], 
\end{equation}
where $a$ is the $SU(N)$ gauge field, and $\mathcal{P}\exp(\cdots)$ is the path-ordered exponential. 
The 't~Hooft loop is a non-genuine line operator, or a defect order parameter, which is defined by the introduction of the non-flat background gauge field:
\begin{equation}
    H(C; \Sigma):= \text{Defect operator that introduces $B_{4\diff}$ with $\diff B_{4\diff}=\frac{2\pi}{N}\delta(C)$}. 
\end{equation}
The 't~Hooft loop cannot be completely specified by the local data of the line $C=\partial \Sigma$, and we need to choose the surface $\Sigma$ that spans it as $B_{4\diff}=\frac{2\pi}{N}\delta(\Sigma)$. 
Equivalently, the 't~Hooft loop lives on the boundary of the $1$-form symmetry generators defined on the open surface $\Sigma$. 
As a consequence, $H(C;\Sigma)$ depends on the surface $\Sigma$ only topologically, which forbids the local counterterm extended along $\Sigma$, and thus its (magnetic) string tension is the well-defined order parameter to diagnose the low-energy quantum phases~\cite{Gukov:2013zka, Kapustin:2013qsa, Kapustin:2013uxa}.
Since we have chosen $k_{\mathrm{UV}}=0$, one may interpret $H(C;\Sigma)$ as the worldline of the fundamental-weight-charge monopoles if we consider the Abelianized regime via adjoint Higgsing, which fits the original definition~\cite{tHooft:1977nqb}.  

For the low-energy topological field theory~\eqref{eq:Z_TQFT}, we can diagnose how the UV line operators flow to the deconfined (non-genuine) lines from the way it couples to $B_{4\diff}$~\cite{Nguyen:2023fun, Maeda:2025ycr}: 
\begin{align}
    W^{n}(C) & \xrightarrow{\text{RG flow}} \exp\left(\im \int_C a_{N/n}\right), \\
    H^{N/n}(C;\Sigma)W^{k}(C) & \xrightarrow{\text{RG flow}} \exp\left(\im \frac{N}{n}\int_{\Sigma}b_{N/n}\right). 
\end{align}
All the deconfined line operators are thus generated by $W^n$ and $H^{N/n}W^k$, and the other lines are confined by showing the area law. 
This is nothing but the Wilson-'t~Hooft classification, which claims the perimeter law for an order-$N$ mutually local subset in the $\mathbb{Z}_N\times \mathbb{Z}_N$ set of the dyonic line operators $\{W^e H^m\}_{e,m\in \mathbb{Z}_N}$. 

\subsection{Review of 3d monopole semiclassics for confinement phases on \texorpdfstring{$\mathbb{R}^3\times S^1$}{R3xS1}}
\label{sec:MonopoleSemiclassics}



Let us explicitly confirm if the Wilson-'t~Hooft classification is actually valid, and then we need an analytically calculable setup of various confinement phases for this purpose. 
As a specific realization, we consider the $3$d monopole semiclassics for the $4$d $SU(N)$ Yang-Mills theory on small $\mathbb{R}^3\times S^1$ with the center-stabilizing double-trace potential for the Polyakov loop~\cite{Unsal:2008ch}. 
The inclusion of the double-trace deformation can be thought of as the effective description of the massive adjoint fermion with the periodic boundary condition~\cite{Unsal:2007jx, Unsal:2007vu, Unsal:2008ch}, and it prevents the confinement-deconfinement phase transition unlike the case of the thermal Yang-Mills theory~\cite{Gross:1980br}. 
If the size $L_4$ of $S^1$ is sufficiently small compared with the strong scale $\Lambda$,
\begin{equation}
    N \Lambda L_4\ll 1, 
\end{equation}
this theory admits the weakly-coupled description of the confinement phases as we will review below. 

It is expected that there exists a smooth path connecting the weakly-coupled confinement regime on small $S^1$ and the strongly-coupled confinement regime on the $\mathbb{R}^4$ limit.\footnote{It is worth noting that the Wilson-'t Hooft classification is a kinematical prediction applicable to any confining phase of a theory with $\mathbb{Z}_N^{[1]}$ symmetry. Therefore, regardless of the adiabatic continuity, observing the dyonic loops within the calculable semiclassical regime is itself a nontrivial task.}
We would like to point out that the lattice numerical study for the double-trace deformed Yang-Mills theory~\cite{Bonati:2018rfg, Bonati:2020lal} observes the qualitative behavior for the topological susceptibility consistent with the adiabatic continuity conjecture. 
Moreover, there exists another semiclassical framework that uses center vortices by considering the $4$d Yang-Mills theory on $\mathbb{R}^2\times T^2$ with the 't~Hooft twisted boundary condition~\cite{Tanizaki:2022ngt, Hayashi:2024qkm}, which has recently been shown to be smoothly connected to the $\mathbb{R}^3\times S^1$ monopole semiclassics \cite{Hayashi:2024yjc, Hayashi:2024psa, Guvendik:2024umd}. 
The lattice numerical study of this $\mathbb{R}^2\times T^2$ setup \cite{Bergner:2025qsm} also observes the behavior of the fundamental string tension consistent with the adiabatic continuity to the confinement state of the $\mathbb{R}^4$ limit. 


\subsubsection{Effective Lagrangian for the 3d monopole semiclassics}

We now review the derivation of the 3d effective theory at small $S^1$ \cite{Unsal:2007jx, Unsal:2007vu, Unsal:2008ch} while keeping the holonomy degrees of freedom.
See also Refs.~\cite{Davies:1999uw, Davies:2000nw} for the case of the $\mathcal{N}=1$ supersymmetric Yang-Mills theory.



On $\mathbb{R}^3\times S^1$, the Polyakov loop, $P_4=\mathcal{P}\exp(\im \int_{S^1} a_4 \diff x_4)$, plays the role of the adjoint Higgs field for the $3$d effective theory, and let us take the Polyakov gauge, which diagonalizes $P_4$ as follows:
\begin{align}
    P_4 = \operatorname{diag}(\rme^{\im \varphi_1}, \cdots, \rme^{\im \varphi_N} )\,,
    \label{eq:PolyakovGauge}
\end{align}
with the constraint $\varphi_1 + \cdots + \varphi_N = 0 ~(\operatorname{mod} 2\pi)$. 
We then parametrize the Polyakov loop by the $N$-component vector field with the constraint, $\Vec{\phi}=(\varphi_1,\ldots, \varphi_{N-1},-\varphi_1-\cdots-\varphi_{N-1})$, and it has the root-vector periodicity:
\begin{align}
    \Vec{\phi} \sim \Vec{\phi} + 2 \pi \vec{\alpha}_i,
\end{align}
where $\vec{\alpha}_i~(i = 1,\cdots,N-1)$ is the positive simple root.\footnote{Our convention for the $SU(N)$ weight and root vectors is the following: Let $\vec{e}_n$ ($n=1,\ldots,N$) be the canonical orthonormal basis of $\mathbb{R}^N$, and we define $\vec{\nu}_n=\vec{e}_n-\frac{1}{N}\sum_{k=1}^{N}\vec{e}_k$, which gives the weight vectors of the defining representation. 
The positive simple roots are given by $\vec{\alpha}_i=\vec{\nu}_i-\vec{\nu}_{i+1}=\vec{e}_i-\vec{e}_{i+1}$ for $i=1,\ldots,N-1$, and the fundamental weights are $\vec{\mu}_i=\vec{\nu}_1+\cdots+\vec{\nu}_i$ also for $i=1,\ldots,N-1$. }
There is no classical potential for $\vec{\phi}$, and the generic point of the classical moduli space gives the adjoint Higgsing, $SU(N)\xrightarrow{\text{Higgs}} U(1)^{N-1}$, since the off-diagonal components of the 3d gluon field ($a^{ij}$ for $i \neq j$) acquire the Kaluza-Klein mass of $\frac{\varphi_i-\varphi_j}{L_4}$ (mod $\frac{2\pi}{L_4}\mathbb{Z}$), while the diagonal components remain massless. 


This resulting 3d $U(1)^{N-1}$ gauge theory can then be reformulated in terms of a $U(1)^{N-1}$-valued scalar field $\vec{\sigma}$, called the dual photon, via 3d Abelian duality, and the dual photon has the weight vector periodicity,
\begin{align}
    \vec{\sigma} \sim \vec{\sigma} + 2 \pi \vec{\mu}_k,
\end{align}
where $\vec{\mu}_k ~(k=1,\cdots, N-1)$ is the fundamental weight.
In summary, the bosonic sector consists of the compact bosons $(\vec{\phi},\vec{\sigma})$ with the target space
\begin{align}
    (\vec{\phi},\vec{\sigma}) \in \frac{\mathbb{R}^{N-1}/2\pi \Lambda_{\mathrm{roots}} \times \mathbb{R}^{N-1}/2\pi \Lambda_{\mathrm{weights}}}{S_N},   \label{eq:target_space_quotient}
\end{align}
where $\Lambda_{\mathrm{roots}}$ and $\Lambda_{\mathrm{weights}}$ are the root and weight lattices, respectively. 
We note that the set of eigenvalues of $P_4$ is gauge invariant while each eigenvalue itself is not physical, and thus there exists the $S_N$ permutation redundancy for $(\vec{\phi},\vec{\sigma})$. 
As a result, we should take the quotient of the target space by $S_N$ as shown in \eqref{eq:target_space_quotient}, and thus we have to deal with the $S_N$ gauge theory. 
One can `fix the $S_N$ gauge redundancy' by restricting the holonomy $\vec{\phi}$ to the fundamental Weyl chamber, e.g.,
\begin{align}
    \vec{\alpha}_i \cdot \vec{\phi} > 0,~~ -\vec{\alpha}_N \cdot \vec{\phi} <2 \pi, \label{eq:Weyl_chambre_gauge_fix}
\end{align}
where $\vec{\alpha}_N = - (\vec{\alpha}_1 + \cdots + \vec{\alpha}_{N-1})$ is the Affine simple root. 
On the boundary of the Weyl chamber, a part of the non-Abelian gauge symmetry is restored and the massless off-diagonal gluons need to be taken into account for the low-energy description.

At the perturbative level, the Polyakov loop effective potential $V_{\mathrm{eff}}(\vec{\phi})$ appears in general. For the pure Yang-Mills case, the Gross-Pisarski-Yaffe (GPY) one-loop potential~\cite{Gross:1980br} prefers the center-broken vacua, $\vec{\phi}=\frac{2\pi}{N}k\vec{\mu}_1$ with $k=0,1,\ldots, N-1$, and thus the small $S^1$ regime is separated from the confinement vacua on $\mathbb{R}^4$ by a phase transition. 
If we introduce several massive adjoint fermions with the periodic boundary condition, the GPY potential flips its sign, and the location of the vacuum becomes center-symmetric, 
\begin{align}
    \vec{\phi}_c = \frac{2 \pi}{N} \vec{\rho} = \frac{2 \pi}{N} (\vec{\mu}_1 + \cdots + \vec{\mu}_{N-1}),
    \label{eq:CenterSymmetricHolonomy}
\end{align}
where $\vec{\rho}$ is called the Weyl vector. 
This center-symmetric holonomy corresponds to $P_4 \propto \operatorname{diag} (1, \rme^{-\frac{2 \pi \im}{N}}, \cdots, \rme^{-\frac{2 \pi \im (N-1)}{N}} )$, which satisfies $\operatorname{tr}P_4^k = 0~(k=1,2,\cdots,N-1)$. 
Let us focus on this center-symmetric situation, and 
the 3d Euclidean Langrangian within the perturbation theory can be written as,
\begin{align}
    \mathcal{L}_{\mathrm{3d}}^{\mathrm{pert.}} = \frac{1}{g^2 L_4} |\diff \vec{\phi}|^2 + \frac{g^2}{16 \pi^2 L_4} \left| \diff \vec{\sigma} + \frac{\theta}{2 \pi } \diff \vec{\phi}  \right|^2 + V_{\mathrm{eff}} (\vec{\phi}), 
\end{align}
where the holonomy potential $V_{\mathrm{eff}}(\vec{\phi})$ prefers the center-symmetric point~\eqref{eq:CenterSymmetricHolonomy}. 

In this setup, the $4$d instanton splits into $N$-types of fundamental monopole-instantons: $N-1$ Bogomol'nyi-Prasad-Sommerfield (BPS) monopoles and one Kaluza-Klein (KK) monopole~\cite{Lee:1997vp, Lee:1998bb, Lee:1998vu, Kraan:1998kp, Kraan:1998pm, Kraan:1998sn}.
Magnetic charges of BPS monopoles are simple roots $\{ \vec{\alpha}_i \}_{i = 1, \cdots , N-1}$, and the magnetic charge of the KK monopole is the affine root $\vec{\alpha}_N$.
For the holonomy in the given Weyl chamber (\ref{eq:Weyl_chambre_gauge_fix}) with the center-symmetric point (\ref{eq:CenterSymmetricHolonomy}), the monopole-instanton vertex $[\mathcal{M}_i]$ and anti-monopole-instanton vertex $[\mathcal{M}_i^*]$ can be written as,
\begin{align}
    [\mathcal{M}_i] &= \zeta_m \rme^{\frac{\im \theta}{N}} \rme^{\im \vec{\alpha}_i \cdot  \left[ \vec{\sigma} + \left( \frac{\theta}{2 \pi} + \frac{4 \pi \im}{g^2} \right) (\vec{\phi} - \vec{\phi}_c) \right]} ~~~(i = 1 , \cdots, N), \notag \\
    [\mathcal{M}_i^*] &= \zeta_m \rme^{- \frac{\im \theta}{N}} \rme^{-\im \vec{\alpha}_i \cdot  \left[ \vec{\sigma} + \left( \frac{\theta}{2 \pi} - \frac{4 \pi \im}{g^2} \right) (\vec{\phi} - \vec{\phi}_c) \right]} ~~~(i = 1 , \cdots, N). \label{eq:monopole-vertex}
\end{align}
with weight $\zeta_m \sim O(\rme^{-\frac{8\pi^2}{Ng^2}})$ that is the monopole fugacity (at the center-symmetric point).
Thus, for holonomy within the Weyl chamber (\ref{eq:Weyl_chambre_gauge_fix}), i.e., when the $S_N$ gauge is fixed, the dilute gas of the monopoles induces the potential
\begin{align}
    V_{\mathrm{monopole}} (\vec{\sigma},\vec{\phi}) = - 2 \zeta_m \sum_{i=1}^N \rme^{- \frac{4 \pi }{g^2} \vec{\alpha}_i \cdot    (\vec{\phi} - \vec{\phi}_c) } &\cos \left( \vec{\alpha}_i \cdot  \left[ \vec{\sigma} + \left( \frac{\theta}{2 \pi} \right) (\vec{\phi} - \vec{\phi}_c) \right] + \frac{\theta}{N}\right) \notag \\
    &~~[\text{for } \vec{\phi} \in \text{Weyl chamber }(\ref{eq:Weyl_chambre_gauge_fix}) ]. \label{eq:Vmono_in_Weylchambre}
\end{align}
This is only defined for the given Weyl chamber, but we can extend the domain of definition of $V_{\mathrm{monopole}} (\vec{\sigma},\vec{\phi})$ through imposing the $S_N$ invariance: $V_{\mathrm{monopole}} (\vec{\sigma},\vec{\phi}) = V_{\mathrm{monopole}} (P(\vec{\sigma},\vec{\phi}))$ for any $P \in S_N$.
By incorporating this monopole potential, we obtain the following effective Lagrangian,
\begin{align}
    \mathcal{L}_{\mathrm{3d}}^{(\vec{\sigma},\vec{\phi})} &= \frac{1}{g^2 L_4} |\diff \vec{\phi}|^2 + \frac{g^2}{16 \pi^2 L_4} \left| \diff \vec{\sigma} + \frac{\theta}{2 \pi } \diff \vec{\phi}  \right|^2 + V_{\mathrm{eff}} (\vec{\phi}) + V_{\mathrm{monopole}} (\vec{\sigma},\vec{\phi}). 
    \label{eq:3dEFT_sigmaphi}
\end{align}
As the holonomy potential $V_{\mathrm{eff}} (\vec{\phi})$ forces the holonomy $\vec{\phi}$ to take the center-symmetric value $\vec{\phi} = \vec{\phi}_c$ already in the perturbative level, we may fix the holonomy degrees of freedom, and have the dual-photon effective theory\footnote{Note the hierarchy of the mass scale: the W-boson has $m_W = \frac{1}{N L}$, the perturbative holonomy potential $V_{\mathrm{eff}} (\vec{\phi})$ typically gives the mass $m_{\phi} \sim \sqrt{N}g/L$, and the dual photon has only a nonperturbative mass.}:
\begin{align}
    \mathcal{L}_{\mathrm{3d}}^{(\vec{\sigma})} &= \frac{g^2}{16 \pi^2 L_4} \left| \diff \vec{\sigma} \right|^2 - 2 \zeta_m \sum_{i=1}^N  \cos \left( \vec{\alpha}_i \cdot \vec{\sigma}  + \frac{\theta}{N}\right). 
    \label{eq:dual_photon_EFT}
\end{align}

\subsubsection{Symmetry in the 3d effective theory}
\label{sec:3dEFTsymmetry}


Here, we review the global symmetry of the $3$d effective theory before explaining the puzzle on the 't~Hooft loops. 
We discuss how the $\mathbb{Z}_N^{[1]}$ symmetry of the Yang-Mills theory is realized in the $3$d monopole semiclassics and also the emergent symmetry that is specific to the effective Lagrangian~\eqref{eq:3dEFT_sigmaphi}. 

The $\left( \mathbb{Z}_N^{[1]} \right)_{\mathrm{4d}}$ symmetry in $4$d Yang-Mills theory is decomposed into the $1$- and $0$-form symmetries in the 3d effective theory on $\mathbb{R}^3\times S^1$;
\begin{align}
    \left( \mathbb{Z}_N^{[1]} \right)_{\mathrm{4d}} \xrightarrow{\text{on }\mathbb{R}^3\times S^1}  \left( \mathbb{Z}_N^{[0]} \right)_{\mathrm{3d}} \times  \left( \mathbb{Z}_N^{[1]} \right)_{\mathrm{3d}}. 
    \label{eq:sym_decomp_4d3d}
\end{align}
Here, the former one $\left( \mathbb{Z}_N^{[0]} \right)_{\mathrm{3d}}$ denotes the center symmetry acting on the Polyakov loop, $P_4\mapsto \rme^{\frac{2\pi \im}{N}}P_4$, and the latter one $\left( \mathbb{Z}_N^{[1]} \right)_{\mathrm{3d}}$ is the 1-form center symmetry acting on the spatial Wilson loop.

First, let us see how this center symmetry $\left( \mathbb{Z}_N^{[0]} \right)_{\mathrm{3d}} \times  \left( \mathbb{Z}_N^{[1]} \right)_{\mathrm{3d}}$ is realized within the 3d effective theory $\mathcal{L}_{\mathrm{3d}}^{(\vec{\sigma},\vec{\phi})}$ in terms of $(\vec{\sigma},\vec{\phi})$.
As $\vec{\phi}$ describes the eigenvalues of the Polyakov loop, one may represent the action of the 0-form center symmetry $\left( \mathbb{Z}_N^{[0]} \right)_{\mathrm{3d}}$ as
\begin{align}
    \left( \mathbb{Z}_N^{[0]} \right)_{\mathrm{3d}} : \vec{\phi} \mapsto \vec{\phi} - 2 \pi \vec{\mu}_1.
    \label{eq:0formCenter_naive}
\end{align}
Although this expression is not manifestly $S_N$-invariant, this action is $S_N$-invariant due to the root-vector periodicity of $\vec{\phi}$.
We should note that this action~\eqref{eq:0formCenter_naive} of the $0$-form center symmetry does not respect the fundamental Weyl chamber~\eqref{eq:Weyl_chambre_gauge_fix}. 
To maintain the constraint~\eqref{eq:Weyl_chambre_gauge_fix}, we need to combine it with the $S_N$-gauge transformation as 
\begin{equation}
    \left( \mathbb{Z}_N^{[0]} \right)_{\mathrm{3d}} : (\vec{\sigma},\vec{\phi}) \mapsto (P_W^{-1} \vec{\sigma}, P_W^{-1}(\vec{\phi} - 2 \pi \vec{\mu}_1)), 
    \label{eq:0formCenter}
\end{equation}
where $P_W$ refers to the cyclic Weyl permutation~\cite{Anber:2015wha}.\footnote{Related to this fact, the monopole-induced potential $V_{\mathrm{monopole}} (\vec{\sigma},\vec{\phi})$ in (\ref{eq:Vmono_in_Weylchambre}) is not symmetric under \eqref{eq:0formCenter_naive}, while it satisfies $V_{\mathrm{monopole}} (\vec{\sigma},\vec{\phi}) = V_{\mathrm{monopole}} (P_W^{-1}\vec{\sigma},P_W^{-1}(\vec{\phi} - 2 \pi \vec{\mu}_1))$ since $P_W \vec{\phi_c} = \vec{\phi_c} - 2 \pi \vec{\mu}_1 $.
This is quite natural because the expression (\ref{eq:Vmono_in_Weylchambre}) is defined only for the fundamental Weyl chamber (\ref{eq:Weyl_chambre_gauge_fix}).
}

The 1-form part $\left( \mathbb{Z}_N^{[1]} \right)_{\mathrm{3d}}$ acts on the spatial Wilson loop, and it is translated to the winding (magnetic) symmetry of the dual photon $\vec{\sigma}$ through the 3d electromagnetic duality.
Hence, the symmetry operator, which is a one-dimensional topological operator (co-dimension-2 operator in 3d), can be written as
\begin{align}
    U_{\left( \mathbb{Z}_N^{[1]} \right)_{\mathrm{3d}}}(C) = \rme^{-\im \vec{\mu}_1\cdot \int_C d\vec{\sigma}}.
\end{align}
Again, this expression is not manifestly $S_N$-invariant but satisfies its invariance due to the periodicity, $\int_C d\vec{\sigma} \in 2 \pi \Lambda_{\mathrm{weights}}$. This reflects the fact that the center symmetry is an invertible symmetry, unlike the non-invertible ones shown below.

Whereas these center symmetries are originally present from the UV theory, the 3d effective theory $\mathcal{L}_{\mathrm{3d}}^{(\vec{\sigma},\vec{\phi})}$(\ref{eq:3dEFT_sigmaphi}) has accidentally enhanced (noninvertible) symmetries \cite{Nguyen:2021yld}.
\begin{itemize}
    \item Electric 1-form symmetry:
    $S_N$ conjugates of $\left( U(1)^{N-1} \right)_{\mathrm{3d, ele}}^{[1]}$ symmetry
\begin{align}
    U_{\vec{\theta} }^{\mathrm{3d, ele} }(C) = \sum_{P \in S_N} P \rme^{\im \vec{\theta} \cdot \int_C d\vec{\sigma}} P^{-1}
\end{align}
with $\vec{\theta} \in \mathbb{R}^{N-1}/(2 \pi \Lambda_{\mathrm{roots}})$.
This symmetry emerges due to the decoupling of the W bosons associated with the adjoint higgsing $SU(N) \rightarrow U(1)^{N-1}$.
    \item Magnetic 1-form symmetry:
    $S_N$ conjugates of $\left( U(1)^{N-1} \right)_{\mathrm{3d, mag}}^{[1]}$ symmetry
\begin{align}
    U_{\vec{\theta} }^{\mathrm{3d, mag} }(C) = \sum_{P \in S_N} P  \rme^{\im \vec{\theta} \cdot \int_C d\vec{\phi}} P^{-1}
\end{align}
with $\vec{\theta} \in \mathbb{R}^{N-1}/(2 \pi \Lambda_{\mathrm{weights}})$.
    \item $\mathrm{Rep}(S_N)^{[1]}$ symmetry.

    In the $S_N$ gauge theory, the $S_N$ Wilson loop is topological due to the flatness of the $S_N$ part.
    This topological operator generates $\mathrm{Rep}(S_N)^{[1]}$ symmetry \cite{Bhardwaj:2017xup}.
    In other words, the emergence of this symmetry represents the decoupling of the ``twist vortex,'' which will be discussed below.
\end{itemize}
When the emergent symmetries are the $0$-form symmetry, we can add the local perturbation (possibly described by a higher-dimensional operator) that violates the emergent symmetry. 
However, since these emergent symmetries are the $1$-form symmetry, any local perturbation does not break it, and we encounter a strong accidental selection rule within the effective theory.

Going back to the $\mathbb{Z}_N$ center symmetry present in the UV, let us add a remark on the coupling to the background gauge field of $\left( \mathbb{Z}_N^{[0]} \right)_{\mathrm{3d}} \times  \left( \mathbb{Z}_N^{[1]} \right)_{\mathrm{3d}}$, denoted by $A_{3\diff}$ and $B_{3\diff}$, respectively. 
Corresponding to the decomposition~\eqref{eq:sym_decomp_4d3d}, these 3d background gauge fields describe the components of the 4d background field $B_{4\diff}$ of $\left( \mathbb{Z}_N^{[1]} \right)_{\mathrm{4d}}$ as follows,
\begin{align}
   B_{\mathrm{4d}} =  A_{\mathrm{3d}} \wedge \frac{dx_4}{L_4} + B_{\mathrm{3d}}. 
\end{align}
We can write the 3d action in the presence of the background gauge field $(A_{\mathrm{3d}}, B_{\mathrm{3d}})$ as~\cite{Tanizaki:2019rbk}
\begin{align}
    S_{\mathrm{3d}}[A_{\mathrm{3d}}, B_{\mathrm{3d}}] &= \int \frac{1}{g^2 L_4} \left|\diff \vec{\phi} + N A_{\mathrm{3d}} \vec{\mu}_1 \right|^2 + \frac{g^2}{16 \pi^2 L_4} \left| \diff \vec{\sigma} + \frac{\theta}{2 \pi } (\diff \vec{\phi} + N A_{\mathrm{3d}} \vec{\mu}_1)  \right|^2 \notag \\
    &~~~~+ \int \diff^3x ~(V_{\mathrm{eff}} (\vec{\phi}) + V_{\mathrm{monopole}} (\vec{\sigma}, \vec{\phi})) + \frac{\im N}{2 \pi } \int \vec{\mu}_1 \cdot \diff \vec{\sigma} \wedge B_{\mathrm{3d}}. \label{eq:backgrd_coupling_explicit}
\end{align}
While this coupling is a natural minimal choice, we have the freedom to modify the action by adding gauge-invariant local terms (counterterms) of the background field:
\begin{align}
    \Delta S_{\mathrm{3d}}[A_{\mathrm{3d}}, B_{\mathrm{3d}}] &= \frac{iNk_{\mathrm{UV}}}{2 \pi } \int A_{\mathrm{3d}} \wedge B_{\mathrm{3d}}~~~~~ (k=0,1,\cdots,N-1),
\end{align}
This ambiguity reflects the dependence on the UV regularization scheme.
Thus, the absolute phase of the partition function depends on the regularization, and only the relative phase (or difference) between theories carries intrinsic physical meaning.
With this in mind, we adopt the canonical choice (\ref{eq:backgrd_coupling_explicit}), i.e. $k_{\mathrm{UV}}=0$, throughout this paper.
This fixes our convention; specifically, the SPT phases discussed hereafter are defined under a UV regularization that is consistent with (\ref{eq:backgrd_coupling_explicit}).
In particular, in this choice, the SPT phase is trivial in the confining vacuum at $\theta = 0$, that is $(\vec{\sigma}, \vec{\phi}) = (\vec{0}, \vec{\phi}_c))$.

As a final preliminary remark before proceeding to the main discussion, we note how the mixed anomaly between $\left( \mathbb{Z}_N^{[1]} \right)_{\mathrm{4d}}$ symmetry and $2\pi$ periodicity of $\theta$ is encoded in the 3d effective theory.
Let us consider the shift: $\theta \mapsto \theta + 2\pi$ in the 3d effective action with background fields (\ref{eq:backgrd_coupling_explicit}).
In order to compensate for the change in the kinetic term, we need to shift the dual photon as\footnote{Remember that $\vec{\sigma}$ has the weight-vector periodicity.
Thus, as a compact scalar, the shift by $N A_{\mathrm{3d}} \vec{\mu}_1$ is possible.
Also, as $\Lambda_{\mathrm{roots}} \subset \Lambda_{\mathrm{weights}}$, the shift $\diff \vec{\sigma} \mapsto \diff \vec{\sigma} - \diff \vec{\phi}$ is well-defined.
},
\begin{align}
    \diff \vec{\sigma} \mapsto \diff \vec{\sigma} - (\diff \vec{\phi} + N A_{\mathrm{3d}} \vec{\mu}_1) 
\end{align}
This redefinition results in a shift in the local counterterm (or SPT phase)\footnote{This anomaly also implies that the ambiguity in the UV regularization scheme corresponds merely to a shift of $\theta$ by $2\pi k$ (for some $k=0,1,2,\cdots, N-1$). }:
\begin{align}
    S_{\mathrm{3d}}[A_{\mathrm{3d}}, B_{\mathrm{3d}}] \xrightarrow{~\theta \mapsto \theta + 2\pi ~} S_{\mathrm{3d}}[A_{\mathrm{3d}}, B_{\mathrm{3d}}] + \frac{\im N}{2 \pi } \int A_{\mathrm{3d}} \wedge B_{\mathrm{3d}},
\end{align}
which is indeed the 3d counterpart of the mixed anomaly between $\left( \mathbb{Z}_N^{[1]} \right)_{\mathrm{4d}}$ symmetry and $\theta$-periodicity~\cite{Gaiotto:2017yup}.

\subsection{Definition of dyonic loops in the 3d monopole semiclassics and the puzzle}


In this section, we describe the definition of the both genuine and non-genuine loop operators in the $3$d monopole semiclassics to be used for the Wilson-'t~Hooft classification. 
It has been well-known that the $3$d monopole semiclassics~\eqref{eq:3dEFT_sigmaphi} shows the area law for the Wilson loops~\cite{Unsal:2007jx, Unsal:2007vu, Unsal:2008ch}, and the nontrivial question that has not been addressed in previous literature is whether there exists the dyonic loop showing the perimeter law. 

We here give the most straightforward definition for the 't~Hooft loop operator, which is also used in Refs.~\cite{Anber:2015wha, Anber:2017rch} to classify the global structure of the $\mathfrak{su}(N)$ Yang-Mills theories, and then we pose a puzzle about the low-energy behaviors of the dyonic operators: 
The dyonic operators in this definition cannot be calculated in the well-defined manner within the $3$d monopole semiclassics, and if one forcibly proceeds their computation, all the dyonic lines turn out to show the area law, which contradicts the expectation from the Wilson-'t~Hooft classification.

\subsubsection{Definition in the 3d monopole semiclassics}

Let us first quickly review the definition of the Wilson loop. Since we have integrated out the off-diagonal gluons, the Wilson loop operator is just described by the Abelian components, and the $3$d Abelian duality implies that it can be expressed as the defect operator that imposes the winding configuration for the dual photon $\sigma$, 
\begin{align}
    W_{\vec{\nu}_j}(C) &=\exp\left(\im \vec{\nu}_j\cdot\int_C \vec{a}\right) \notag\\
    &= \text{Defect requiring $\int_{S^1}\diff \vec{\sigma}=2\pi \vec{\nu}_j$ for small $S^1$ around $C$. }
\end{align}
Since the monopole potential prefers the specific location for the dual photon $\vec{\sigma}$ as its vacua, we can show the area law as in the case of the Polyakov mechanism~\cite{Unsal:2007jx, Unsal:2007vu, Unsal:2008ch}. 
More interestingly, when we create the domain wall connecting the confining vacua at $\theta=0$ and $\theta=2\pi$, the Wilson loop on the wall can be shown to be deconfined~\cite{Sulejmanpasic:2016uwq, Cox:2019aji}. 
Later, the deconfinement on the wall gets an interpretation as the kinematic consequence of the anomaly inflow, which comes out of the fact that these confining vacua belong to different SPT states with the $\mathbb{Z}_N^{[1]}$ symmetry~\cite{Gaiotto:2017yup, Komargodski:2017smk}. 


Next, let us discuss the definition of the 't Hooft loop, which is a non-genuine line operator and characterized as the boundary of the $\mathbb{Z}_N^{[1]}$ center-symmetry defect on some open surface. 
Since the 4d 1-form symmetry splits into 3d 0-form and 1-form symmetries after the $S^1$ compactification as discussed in \eqref{eq:sym_decomp_4d3d}, 
we have two types of the 't~Hooft operators on $\mathbb{R}^3\times S^1$:
\begin{itemize}
    \item Spatial 't Hooft loop: boundary of the $\left( \mathbb{Z}_N^{[0]} \right)_{\mathrm{3d}}$ symmetry defect,
    \item Temporal 't Hooft loop: boundary of the $\left( \mathbb{Z}_N^{[1]} \right)_{\mathrm{3d}}$ symmetry defect.
\end{itemize}

In the 3d effective theory with the holonomy and dual photon, we define the spatial 't~Hooft loop as
\begin{align}
    H_{\mathrm{IR}}(C;\Sigma) :=  \sum_{j=1}^N U_{\Delta\vec{\phi} = -2 \pi \vec{\nu}_j}(\Sigma), 
    \label{eq:spatial_tHt_def}
\end{align}
where $U_{\Delta\vec{\phi} = -2 \pi \vec{\nu}_j}(\Sigma)$ is the defect that shifts the holonomy as $\vec{\phi} \mapsto \vec{\phi} -2 \pi \vec{\nu}_j$ on the open surface $\Sigma$, and $\vec{\nu}_j$ is the weight vector of the fundamental representation.
Note that the sum over the $\vec{\nu}_j$ is necessary for the permutation invariance.
These vectors $\{ \vec{\nu}_1, \vec{\nu}_2, \cdots, \vec{\nu}_N \}$ are generated by permuting the fundamental weight $\vec{\mu}_1$. For the derivation of this expression from the canonical operator formalism, see Appendices of Ref.~\cite{Anber:2015wha}.

We have to add a remark about $U_{\Delta\vec{\phi} = -2 \pi \vec{\nu}_j}(\Sigma)$.
To define the open surface defect, we must carefully consider the lift of the field $\vec{\phi}$ from $\mathbb{R}^{N-1}/(2 \pi \Lambda_{\mathrm{roots}})$ to $\mathbb{R}^{N-1}$.
Specifically, we are left with the freedom to determine the monodromy of $\vec{\phi}$ around the boundary loop $C$.
In our definition, we choose the following specification for $U_{\Delta\vec{\phi}}(\Sigma)$: for a loop $C'$ winding around $C = \partial \Sigma$,
\begin{align}
    \int_{C'} \diff \vec{\phi} + \Delta \vec{\phi} = 0.
\end{align}
In general, one may choose any root vector $2 \pi \vec{\alpha}$ as the right-hand side.
However, as we will shortly see, this modification corresponds to attaching a genuine-line 't Hooft loop of the magnetic charge $\vec{\alpha}$.
Thus, we choose the simplest assignment.

To justify this definition, let us mention several reasons to think the definition (\ref{eq:spatial_tHt_def}) is a natural one:
\begin{itemize}
    \item The operator $U_{\Delta\vec{\phi} = -2 \pi \vec{\nu}_j}(\Sigma)$ imposes the monodromy in $\vec{\phi}$, which is equivalent to
\begin{align}
    \int_{C'\times S^1}\diff \vec{a} = 2 \pi \vec{\nu}_j,
\end{align}
in terms of the Abelian components of the 4d gauge field $\vec{a}$.
Here, $C'$ is a contour winding around the loop $C=\partial \Sigma$.
Therefore, this defect describes the worldline of the magnetic monopole with the weight charge $2\pi\vec{\nu}_j$, which is the standard definition of the 't~Hooft loop in the Abelian gauge theory. 

    \item As we shall show in details in Appendix \ref{app:Lattice}, we can derive the expression~\eqref{eq:spatial_tHt_def} starting from the standard definition of the 't~Hooft loop in the lattice gauge theory. 
    Here, let us briefly summarize the idea for the derivation. 

    After dimensional reduction along the $S^1$ direction, the 't Hooft loop becomes the $\mathbb{Z}_N$-twist for the hopping term of the Polyakov loop $P_4$ on $\Sigma$.
    Roughly speaking, this corresponds to\footnote{In the standard Wilson lattice formulation, the winding number cannot be strictly specified. Consequently, a lattice operator corresponds to a superposition of continuum operators with different windings.} $H_{\mathrm{lat}}(C;\Sigma) \sim \sum_{\vec{\alpha}\in \Lambda_{\mathrm{roots}}} U_{\Delta\vec{\phi} = -2 \pi \vec{\mu}_1 + 2 \pi \vec{\alpha} }(\Sigma)$.
    By extracting minimal terms among $-2 \pi \vec{\mu}_1 + 2 \pi \vec{\alpha} $, we recover our definition (\ref{eq:spatial_tHt_def}). 

\end{itemize}

Similarly, the non-genuine temporal 't Hooft loop $H(\{p,p'\};\gamma)$ is defined on the open line $\gamma$, with $\partial \gamma = \{p,p'\}$ in the 3d language.
This 't Hooft loop $H(\{p,p'\};\gamma)$ is characterized by the $\left( \mathbb{Z}_N^{[1]} \right)_{\mathrm{3d}}$ defect on an open line.
Recall that the $\left( \mathbb{Z}_N^{[1]} \right)_{\mathrm{3d}}$ defect is 
\begin{align}
    \sum_{j=1}^N \rme^{\im \vec{\nu}_j\cdot \int_{C} d\vec{\sigma}}
\end{align}
for a loop $C$, because the Wilson loop is the monodromy defect of the dual photon from the 3d electromagnetic duality.
The sum over $\{ \vec{\nu}_1, \vec{\nu}_2, \cdots, \vec{\nu}_N \}$ is taken due to the manifest permutation invariance.
As above, we will construct an operator like 
\begin{align}
       `` \left[ \left( \sum_{j=1}^{N} \rme^{-\im  \vec{\nu}_{j} \cdot \vec{\sigma}(p)}\right) \times \left(\mathbb{Z}_N\text{ topological line on } \gamma \right) \times \left( \sum_{j'=1}^{N} \rme^{\im  \vec{\nu}_{j} \cdot \vec{\sigma}(p')}\right) \right]",
\end{align}
which is not a precise expression since $\vec{\sigma}$ has the weight-vector periodicity.
A proper expression can be given by,
\begin{align}
     H(\{p,p'\};\gamma) := \sum_{j,j'=1}^{N}  \rme^{\im  \vec{\alpha}_{j',j} \cdot \vec{\sigma}(p)} \rme^{\im \int_\gamma \vec{\nu}_j \cdot d \vec{\sigma}}  \label{eq:def_temp_tHloop}
\end{align}
with $\vec{\alpha}_{j',j} = \vec{e}_{j'} - \vec{e}_{j}$.
Indeed, this definition can arise from the lattice observation, see Appendix \ref{app:lat_temp_tHloop} for details.
Also, as the genuine-line 't Hooft loop of a root-vector magnetic charge $\vec{\alpha}$ is $\rme^{\im\vec{\alpha}\cdot \vec{\sigma}}$, the above 't Hooft loop is the natural one as ``the 't Hooft loop of a weight-vector magnetic charge'' in the abelianized gauge theory.

It is noteworthy that the summations over the permutations are performed independently at the two points $\{p,p'\}$ in the temporal 't Hooft loop (\ref{eq:def_temp_tHloop}).
Let us remark on this aspect from the perspective of the dual-photon effective theory (\ref{eq:dual_photon_EFT}).
At first glance, after fixing the holonomy, the operator $\rme^{\im \int_\gamma \vec{\nu}_j \cdot d \vec{\sigma}}$ (for some $j$) appears to be a natural candidate for the temporal 't Hooft loop.
Nevertheless, it is not appropriate to identify this operator with the 't Hooft loop.
For the purpose of the Wilson-'t Hooft classification, it is crucial to ensure that the operator carries the proper $\left( \mathbb{Z}_N^{[0]} \right)_{\mathrm{3d}}$ charge assignments, i.e., no (electric) $\left( \mathbb{Z}_N^{[0]} \right)_{\mathrm{3d}}$ charge on the boundary for 't Hooft loop. 
On the other hand, the dual-photon effective theory is obtained after fixing the holonomy field $\vec{\phi}$ to the center-symmetric point $\vec{\phi} = \vec{\phi}_c$ (fixing $S_N$ gauge at the same time).
Due to the fixing procedure, the dual-photon effective theory implicitly involves the holonomy field.
Indeed, in the dual-photon effective theory, the center symmetry $\left( \mathbb{Z}_N^{[0]} \right)_{\mathrm{3d}}$ acts as the cyclic Weyl permutation\footnote{This is not a gauge redundancy, as this Weyl permutation only acts on the dual photon.}, which compensates for the original transformation $\vec{\phi} \mapsto \vec{\phi} - 2\pi \vec{\mu}_1$.
To achieve the proper charge assignment, the boundary of the non-genuine 't Hooft loop should be invariant under the $\left( \mathbb{Z}_N^{[0]} \right)_{\mathrm{3d}}$ symmetry.
Consequently, at both endpoints $\{p, p'\}$, we impose manifest cyclic permutation invariance for the dual photon\footnote{
Note that each $\{p, p'\}$ should be neutral under the $\left( \mathbb{Z}_N^{[0]} \right)_{\mathrm{3d}}$ symmetry.
For exmaple, although $\sum_{j=1}^N \rme^{\im \int_\gamma \vec{\nu}_j \cdot d \vec{\sigma}}$ is invariant under the $\left( \mathbb{Z}_N^{[0]} \right)_{\mathrm{3d}}$ transformation as a total, the transformation at one point, $p$ or $p'$, is nontrivial.
}.

\subsubsection{Puzzle about the spatial dyonic loops}

As we have found the definition of the 't~Hooft loop in terms of $(\vec{\sigma},\vec{\phi})$, it seems that we can now compute the expectation values of the dyonic loop operators to judge the SPT levels using the Wilson-'t~Hooft classification. 
However, this is not the case: 
The definition~\eqref{eq:spatial_tHt_def} of the spatial 't Hooft loop $H_{IR}(C; \Sigma)$  only acts on the holonomy $\vec{\phi}$, so it does not work in the dual photon effective theory $\mathcal{L}_{\mathrm{3d}}^{(\vec{\sigma})}$ after integrating out the holonomy field $\vec{\phi}$.
Moreover, even if we use the 3d effective theory $\mathcal{L}_{\mathrm{3d}}^{(\vec{\sigma},\vec{\phi})}$ with both the dual photon and holonomy fields, 
the winding-number constraint on the holonomy $\vec{\phi}$ requires that we must deal with the boundary of the fundamental Weyl chamber~\eqref{eq:Weyl_chambre_gauge_fix}, where the Abelianization fails.\footnote{For example, in the $SU(2)$ case, the holonomy can be parameterized by one $2\pi$-periodic compact scalar $\phi$, and the center symmetric points are $\phi = \pm \pi/2~(\operatorname{mod}2\pi)$. The 3d effective theory becomes singular at $\phi = 0, \pi $, where $P_4 = \pm 1$, and the Abelianization, $SU(2)\xrightarrow{\mathrm{Higgs}}U(1)$, does not occur at these points. } 
Thus, the $3$d low-energy effective theory is insufficient to compute the expectation values of the dyonic loop operators, $H_{\mathrm{IR}}W^k$.

It is still an interesting question to ask what we would get if we forcibly proceed the computation of the dyonic loop operators, $H_{\mathrm{IR}}W^k$, with neglecting those singularities. 
Due to the presence of the holonomy potential $V_{\mathrm{eff}}(\vec{\phi})$ to set $\vec{\phi}=\vec{\phi}_c$ at the vacuum, the above definition of $H_{IR}(C; \Sigma)$ would give the area-law falloff even in the confining phase since it requires the kink configuration of $\vec{\phi}$. 
The same consequence is true for any dyonic loops $H_{\mathrm{IR}}W^k$, but this is not the expected behavior from the Wilson-'t Hooft classification.

In short,  the 't Hooft loop is not fully described within the framework of the 3d effective theory, nor does it exhibit the behavior expected from the Wilson-'t Hooft classification. 
This is the main problem we will resolve in what follows by giving the refined definition of the 't~Hooft loop operators.
Furthermore, we will show that the Wilson-'t Hooft classification works with the new 't Hooft loop.

\section{Twist vortices and the screening of the 't Hooft loop}
\label{sec:Screening_tHooftloop}

For a solution to the above problem, the key notion is the screening of the loop operators.
In the bulk, we write a 3d low-energy effective theory after integrating out heavy off-diagonal degrees of freedom, and the process of the renormalization group (RG) flow can be summarized as follows:
\begin{align}
    &\text{UV theory: }(\text{4d}~SU(N)~ \text{Yang-Mills theory on }\mathbb{R}^3\times S^1) \notag \\
    &~~~~~ \Downarrow~(\text{Integrating out heavy degrees of freedom}) \notag \\
    &\text{IR theory }\mathcal{L}_{\mathrm{3d}}^{(\vec{\sigma},\vec{\phi})} :~
    (\text{3d}~U(1)^{N-1}\rtimes S_N~ \text{gauge theory }+U(1)^{N-1}~\text{-valued holonomy on }\mathbb{R}^3) \notag\\
    &~~~~~~~~~~~~~~~~~~~~~~~~ \equiv(\text{3d $[U(1)^{N-1}\times U(1)^{N-1}]/S_N$ compact boson on $\mathbb{R}^3$})
    \notag \\
    &~~~~~ \Downarrow~(\text{Set the holonomy at the center-symmetric minimum of }V_{\mathrm{eff}} (\vec{\phi})) \notag \\
    &\text{IR theory' }\mathcal{L}_{\mathrm{3d}}^{(\vec{\sigma})} :~(\text{3d}~U(1)^{N-1}~\text{compact boson on }\mathbb{R}^3)     
\end{align}
The last step is optional, and we mainly revisit the discussion of the RG flow to integrate out the off-diagonal gluons from the UV theory to the IR theory $\mathcal{L}_{\mathrm{3d}}^{(\vec{\sigma},\vec{\phi})}$ in this section.
Under this RG flow, the extended objects may not evolve in a straightforward manner and undergo screening by other line operators that are dynamical in the UV theory.

We will show that, after this screening process, the screened 't Hooft loop operator $\left(H(C; \Sigma) \right)_{\mathrm{screened}}$ properly works within the dual-photon effective theory $\mathcal{L}_{\mathrm{3d}}^{(\vec{\sigma})} $.

\subsection{Example: Screening of the Wilson loops with the trivial \texorpdfstring{$N$}{N}-ality}
\label{sec:Ex_screening_charge-N}

Let us begin with the well-known example, in which the screening of the loop operators plays the crucial role. 
The charge-$N$ Wilson loop (i.e., the $N$-th power of the fundamental Wilson loop $W^N(C)$) shows the perimeter-law falloff in the UV theory since the confining string can be broken via soft gluon exchange. 
However, the charge-$N$ Wilson loop obeys the area law in the $3$d effective theory $\mathcal{L}_{\mathrm{3d}}^{(\vec{\sigma},\vec{\phi})}$ and the string breaking is forbidden due to the absence of off-diagonal gluons. 

To resolve this discrepancy, we need to reinstate heavy off-diagonal gluons, i.e., $W$-bosons with root-vector electric charges for the computation of the loop operators. 
For the computation of the local operators in the bulk, we can simply integrate out these heavy $W$-bosons. 
However, in this process, the extended object (Wilson loop) may evolve in the following way: schematically,
\begin{align}
    W_{\mathrm{UV}}^N(C) \longrightarrow W_{\mathrm{IR}}^N(C) + \sharp \sum_{\vec{\alpha}: \mathrm{roots}}\rme^{-\sharp |C|} W_{\mathrm{IR}}^N(C) W_{\mathrm{IR}}^{\vec{\alpha}}(C)+ \cdots.
\end{align}
In general, we should sum up all possible worldlines of $W$-bosons, but for the sake of presentation, we showed only characteristic terms.

Since the $W$-bosons are heavy, these corrections have a strong suppression factor $\rme^{-\sharp |C|}$.
No matter how small the suppression factor is, this is at most a perimeter-law suppression. 
Since $W_{\mathrm{IR}}^N(C)$ obeys the area law within the $3$d effective theory, if some of $W^N_{\mathrm{IR}}W^{\vec{\alpha}}_{\mathrm{IR}}$ obey the perimeter law, the dominant contribution for $W^N_{\mathrm{UV}}(C)$ comes from those $W$-boson-screened Wilson loops when the loop $C$ becomes asymptotically large enough. 
This is indeed the case since we can show that some of these corrections, arising from the worldlines of $W$-bosons, obey the perimeter-law falloff.
Hence, for the large Wilson loop, the correct IR object is just the trivial operator in the IR effective theory, after the perimeter-law factor $\rme^{-\sharp |C|}$ is renormalized: $W_{\mathrm{UV}}^N(C) \rightarrow  \left( W^N(C) \right)_{\mathrm{screened}} = 1$.

\subsection{Twist vortex in the \texorpdfstring{$S_N$}{SN} gauge theory}

The situation for the 't Hooft loop is quite parallel to the screening of the charge-$N$ Wilson loop.
The key object here is the ``twist vortex'' of the $S_N$ gauge theory,\footnote{In the context of $O(2) = U(1)\rtimes S_2$ gauge theory, the twist vortex is also known as Alice string or Cheshire string \cite{Schwarz:1982ec, Alford:1990mk, Bucher:1991qhl}. 
See also \cite{Jacobson:2024muj} for its construction on the modified Villain lattice.} and let us discuss this object in this subsection. 
Within the $S_N$ gauge theory, the twist vortex is treated as an operator instead of the dynamical object, but it is a heavy but dynamical object in the full UV theory. 
The screening of 't~Hooft loops by the dynamical twist vortices shall be discussed in the next subsection.

The twist vortex of the $S_N$ gauge theory is a co-dimension-2 defect imposing the conjugacy class of the holonomy around the defect. 
It is important to note that we cannot specify the holonomy itself since the holonomy in the non-Abelian gauge theory transforms in the adjoint representation under the gauge transformation. 
The twist vortex operator is gauge invariant since we have only specified the gauge equivalence class of the holonomy using the conjugacy class, $[\sigma]=\{\tau\sigma\tau^{-1}\,|\, \tau\in S_N\}$.\footnote{
The conjugacy classes of $S_N$ are totally determined by ``cycle types'': Any permutations can be written as the product of the cyclic permutations (which is called the cycle decomposition), and the cycle type is the numerical data that gives the number of cycles of each size in the cycle decomposition. If two permutations have the same cycle types, then they are conjugate with each other, and vice versa. }
Among them, we focus on one class $[P_W]$ that includes the cyclic Weyl permutation $P_W$: $(\vec{\sigma}, \vec{\phi}) \mapsto (P_W \vec{\sigma}, P_W \vec{\phi})$, as it turns out to play a pivotal role for the screened 't~Hooft loop.

When the loop $C$ has a surface $\Sigma$ such that $\partial \Sigma = C$, one may express the twist vortex $T_{\mathrm{twist}}(C)$ corresponding to $[P_W]$ as
\begin{align}
    T_{\mathrm{twist}}(C) = \sum_{\sigma \in [P_W]} U_{\sigma}(\Sigma), \label{eq:twisted_vortex_expression}
\end{align}
where $U_{\sigma}(\Sigma)$ denotes the permutation transformation $\sigma$ on the open surface $\Sigma$. 
Let us here emphasize that the twist vortex $T_{\mathrm{twist}}(C)$ is the genuine loop operator in $3$d, and the above expression~\eqref{eq:twisted_vortex_expression} is its convenient formula for the later purpose especially when $C$ is a boundary of some open surface. 

For the case of our interest, the $3$d gauge theory before taking the Abelian duality has the gauge group $U(1)^{N-1}\rtimes S_N$. 
Then, the definition of the twist vortex requires the extra integration over the $U(1)^{N-1}$ part, and the expression~\eqref{eq:twisted_vortex_expression} becomes 
\begin{equation}
    T_{\mathrm{twist}}(C)=\int_{U(1)^{N-1}}\diff \tilde{h}\sum_{\sigma \in [P_W]} \tilde{U}_{\tilde{h}^{-1}\sigma \tilde{h}}(\Sigma), 
\end{equation}
where $\tilde{U}_{\tilde{h}^{-1}\sigma \tilde{h}}(\Sigma)$ is the $U(1)^{N-1}\rtimes S_N$ transformation on the open surface $\Sigma$. 

\subsection{Screening of 't Hooft loop by twist vortex}

In the $3$d IR effective theory $\mathcal{L}_{\mathrm{3d}}^{(\vec{\sigma},\vec{\phi})}$, the twist vortices are non-dynamical: The $S_N$ gauge field is flat and merely describes the redundancy of the compact-boson description $(\vec{\sigma}, \vec{\phi})$ as shown in (\ref{eq:target_space_quotient}).
On the other hand, in the full UV description, the $U(1)^{N-1}\rtimes S_N$ group is a part of the $SU(N)$ gauge group, and thus the twist vortex should be treated as a dynamical object even though it is possibly a heavy object. 
In Appendix~\ref{app:Lattice}, we explicitly construct a gauge-field configuration corresponding to the twist vortex in a simplified lattice setup. 

The fact that twist vortices are dynamical in the UV theory tells us that the IR loop operators may be dressed by those twist vortices compared with the naive reduction from the UV expression under the RG flow. 
Now, we can propose a resolution of the puzzle about the spatial 't Hooft loop by considering the screened loop operator,
\begin{align}
   \left(H(C; \Sigma) \right)_{\mathrm{screened}} = H_{\mathrm{IR}}(C; \Sigma) T_{\mathrm{twist}}(C). \label{eq:def_screened_tH_loop}
\end{align}
In principle, we should consider any kinds of the screening effect for $H_{\mathrm{IR}}$ by all possible line operators that are dynamical in the UV theory, but it turns out that the above one is sufficient to obtain the correct low-energy behavior consistent with the Wilson-'t~Hooft classification. 

As a first step, let us show that the above screened 't Hooft loop $\left(H(C; \Sigma) \right)_{\mathrm{screened}}$ is described in a well-defined manner within the $3$d dual-photon effective theory $\mathcal{L}_{\mathrm{3d}}^{(\vec{\sigma})} $.
To obtain this effective theory, the $S_N$ gauge redundancy is fixed by choosing a center symmetric point (a minimum of $V_{\mathrm{eff}}(\vec{\phi})$), e.g., $P_4 = C^{-1}$, that is $\vec{\phi} = \vec{\phi}_c = \frac{2 \pi}{N} \vec{\rho}$.
Among terms in $\left(H(C; \Sigma) \right)_{\mathrm{screened}} = H_{\mathrm{IR}}(C; \Sigma) T_{\mathrm{twist}}(C)$, there is a term which does not shift the holonomy from the center-symmetric point $\vec{\phi} = \vec{\phi}_c$.
In our choice $\vec{\phi}_c = \frac{2 \pi}{N} \vec{\rho}$, as the Weyl vector satisfies $P_W^{-1} \vec{\rho} = \vec{\rho} + N (P_W^{-1}\vec{\mu}_1)= \vec{\rho} + N \vec{\nu}_N$ (on the lift in $\mathbb{R}^{N-1}$), the inverse cyclic permutation $P_W^{-1}$ in $T_{\mathrm{twist}}(C)$ compensates for the center transformation for $\vec{\phi}$ by $H_{\mathrm{IR}}(C;\Sigma)$.
By extracting such a term preserving $\vec{\phi} = \vec{\phi}_c$, we define
\begin{align}
    H^{(\vec{\sigma})}(C; \Sigma) &: =  \left. \left(H(C; \Sigma) \right)_{\mathrm{screened}}\right|_{\vec{\phi} = \vec{\phi}_c~\mathrm{fixed}} \notag\\ 
    &= U_{P_W^{-1}}^{(\vec{\sigma})}(\Sigma), 
    \label{eq:tH_loop_dualphotonEFT}
\end{align}
where $U_{P_W^{-1}}^{(\vec{\sigma})}(\Sigma)$ is the cyclic Weyl permutation only on the dual photon $\vec{\sigma}$ on the open surface $\Sigma$. 
For this operator, the holonomy does not touch the boundary of the Weyl chamber, and thus the computation within the Abelian effective theory is totally well defined. 

To express $U_{P_W^{-1}}^{(\vec{\sigma})}(\Sigma)$ more precisely, we need to carefully treat a lift of the dual photon $\vec{\sigma}$ from $\mathbb{R}^{N-1}/2 \pi \Lambda_{\mathrm{weights}}$ to $\mathbb{R}^{N-1}$.
Let $\tilde{U}_{P_W^{-1}}^{(\vec{\sigma})}(\Sigma)$ be a cyclic Weyl permutation operator for the $\mathbb{R}^{N-1}$-valued lift $\vec{\tilde{\sigma}}$.
By changing a lift $\vec{\tilde{\sigma}} \mapsto \vec{\tilde{\sigma}} + 2\pi \vec{\mu}$ with $\vec{\mu} \in \Lambda_{\mathrm{weights}}$, this defect is subject to the change $\tilde{U}_{P_W^{-1}}^{(\vec{\sigma})}(\Sigma) \mapsto \tilde{U}_{P_W^{-1}}^{(\vec{\sigma})}(\Sigma) W_{P_W^{-1} \vec{\mu} - \vec{\mu}} (C)$, because the Wilson loop is the monodromy defect of the dual photon $\vec{\sigma}$.
Note that any root charge can be generated as $\{ P_W^{-1} \vec{\mu} - \vec{\mu}| \vec{\mu} \in \Lambda_{\mathrm{weights}}\} = \Lambda_{\mathrm{roots}}$.
It would be natural to sum up all the possible choices of the lift, and then the explicit expression using a specific lift $\tilde{\sigma}$ is given by
\begin{align}
    U_{P_W^{-1}}^{(\vec{\sigma})}(\Sigma) = \sum_{\vec{\alpha} \in \Lambda_{\mathrm{roots}}}\tilde{U}_{P_W^{-1}}^{(\vec{\sigma})}(\Sigma) W_{\vec{\alpha}} (C). 
    \label{eq:tH_loop_dualphotonEFT_lifted}
\end{align}
Note that the absorption of the root charge Wilson loop is also manifest in terms of the $U(1)^{N-1}$ gauge theory before taking the electromagnetic duality.
Indeed, as $\{ P_W^{-1} \vec{\mu} - \vec{\mu}~|~ \vec{\mu} \in \Lambda_{\mathrm{weights}}\} = \Lambda_{\mathrm{roots}}$, any root charge Wilson loop can be written as $W_{\vec{\alpha}}(C) = \rme^{\im (P_W^{-1} \vec{\mu} - \vec{\mu})\cdot \int_{\Sigma} \diff \vec{a}}$.
Such an operator is absorbed by $U_{P_W^{-1}}^{(\vec{\sigma})}(\Sigma) $, which is the Weyl permutation defect for the $U(1)^{N-1}$ gauge sector.

\subsection{Perspective from emergent symmetry}
\label{sec:emergent_symmetry}

In this section, we have emphasized that the effect of screening is crucial for correctly evaluating the long-distance behavior of extended operators. Here, we comment on this screening mechanism from the perspective of symmetries.

The necessity of considering screening arises when the behavior of a loop operator differs between the original UV theory and the IR effective theory. Typically, this discrepancy occurs when charged matter fields responsible for screening are heavy and are integrated out in the low-energy effective theory. 
This phenomenon corresponds to an emergence of new symmetry in the IR theory that was not present in the original UV theory, which is so-called emergent symmetry.

Consider the example of the charge-$N$ Wilson loop $W^N (C)$, which has trivial $N$-ality.
As mentioned in Section \ref{sec:Ex_screening_charge-N}, the charge-$N$ Wilson loop obeys the perimeter law in the original UV $SU(N)$ Yang-Mills theory. 
However, the W-bosons are decoupled in the abelianized IR theory. As a result, the IR theory acquires an emergent electric $U(1)^{N-1}$-like 1-form symmetry (which is technically a noninvertible symmetry, see Section \ref{sec:3dEFTsymmetry}), which would yield a stricter selection rule than the original $\mathbb{Z}_N^{[1]}$ symmetry. 
If one naively evaluates the Wilson loop within the IR abelianized effective theory, it leads to an area law, contradicting the true long-distance behavior in the original theory.
Hence, in this example, the decoupling of the W-bosons, which causes the accidental area law of the charge-$N$ Wilson loop, can be rephrased as the enhancement of the 1-form symmetry in the IR effective theory.

This section claims that a similar logic applies to the 't Hooft loop.
The IR effective theory possesses the emergent $\operatorname{Rep}(S_N)^{[1]}$ symmetry, that represents the decoupling of twist vortices.
In the IR effective theory, this emergent symmetry yields an accidental selection rule. 
For instance, the naive 't Hooft loop $H_{IR}(C;\Sigma)$ and the composite operator $H_{IR}(C;\Sigma) T_{\text{twist}}(C)$ are distinguished as distinct operators in the IR effective theory, as they carry different charges under the emergent symmetry. 
From this viewpoint, the area law of the naive 't Hooft loop $H_{IR}(C;\Sigma)$  is a consequence of this accidental selection rule.

By definition, this emergent symmetry does not exist in the original UV theory. Consequently, the loop operator in the UV theory should be understood as a superposition of all operators in the IR effective theory that are distinct only due to this emergent symmetry. 
For instance, since the emergent $\operatorname{Rep}(S_N)^{[1]}$ symmetry is absent in the UV theory, the original 't Hooft loop $H_{\text{UV}}(C;\Sigma) $ includes both the naive loop $H_{\text{IR}}(C;\Sigma) $ and the twist-vortex-attached loop $H_{\text{IR}}(C;\Sigma)  T_{\text{twist}}(C)$ in the IR description. 
Therefore, to correctly reproduce the long-distance behavior of the extended object by using the IR effective theory, one must take into account all possible IR operators with which the original operator mixes.
If at least one of the operators obeys the perimeter law, the original operator in the UV theory should be regarded as obeying the perimeter law.
This is a kinematical account of the screening mechanism.

In other words, the loop operators in the IR effective theory $H_{\text{IR}}(C)$ and $H_{\text{IR}}(C) T_{\text{twist}}(C)$ have the same UV quantum number, and thus we can utilize $H_{\text{IR}}(C) T_{\text{twist}}(C)$ for the Wilson-'t Hooft classification.

\section{Behavior of dyonic loops in confining phases}
\label{sec:behavior_dyonicloops}

Now, we are fully prepared to investigate the behavior of line operators predicted by the Wilson-'t Hooft classification.
In the 3d dual photon effective theory (\ref{eq:dual_photon_EFT}), there are $N$ confining phases, depending on $\theta$: the vacuum configuration is,
\begin{align}
    \vec{\sigma} = \vec{\sigma}_k := - \frac{2 \pi k}{N} \vec{\rho}~~~~(\mathrm{for}~|\theta - 2 \pi k| < \pi), \label{eq:dual_photon_vacuum}
\end{align}
with $k = 0,1,\cdots, N-1 ~(\operatorname{mod}N)$.

\subsection{Spatial loops}
As the dual photon is subject to the monopole potential, the Wilson loop exhibits the area law in all $N$ vacua $\vec{\sigma} = \vec{\sigma}_k$: for any weight vector $\vec{\mu} \in \Lambda_{\mathrm{weights}} \setminus \{ 0\}$,
\begin{align}
    \braket{W_{\vec{\mu}} (C) }_{\vec{\sigma} = \vec{\sigma}_k} = (\text{area-law terms}).
\end{align}
Thus, the system is in the confinement phase, where the $\mathbb{Z}_N^{[1]}$ symmetry is unbroken.

Let us compute the screened 't~Hooft loop $H^{(\vec{\sigma})}(C;\Sigma)$ defined by \eqref{eq:tH_loop_dualphotonEFT}. 
First, we consider the $k=0$ vacuum: $\vec{\sigma} = 0$.
Since this vacuum is manifestly invariant under $\vec{\sigma}\mapsto P_W^{-1}\vec{\sigma}$ on the nose (i.e. without using any periodicity), the twisted boundary condition across $\Sigma$ does not require the appearance of the kink configuration, and thus the leading behavior is given by the perimeter law,
\begin{align}
    \braket{H^{(\vec{\sigma})}(C; \Sigma)}_{\vec{\sigma} = 0} = 1 + (\text{area-law terms}).
\end{align}
This suggests that the $k=0$ vacuum on $\mathbb{R}^3\times S^1$ is smoothly connected to the monopole-condensing confinement vacuum on the $\mathbb{R}^4$ limit rather than the dyon-condensing ones\footnote{
At small $S^1$, the mass gap is generated by the Coulomb gas of monopole instantons rather than the $4$d monopole/dyon condensation \cite{Poppitz:2011wy, Anber:2017rch}.
The adiabatic continuity claims that these two pictures are continuously connected through a crossover.
Note also that we use the (non-genuine) dyonic loops as the order parameter, and its perimeter law derived here does not arise from the monopole/dyon condensation.
}. 

Next, we consider another vacuum $\vec{\sigma} = \vec{\sigma}_k~(k \neq 0)$.
From $P_W^{-1} \vec{\sigma}_k = \vec{\sigma}_k - 2 \pi k \vec{\nu}_N$, all terms in (\ref{eq:tH_loop_dualphotonEFT_lifted}) obey the area-law falloff because the term $\tilde{U}_{P_W^{-1}}^{(\vec{\sigma})}(\Sigma)$ requires a kink of $\vec{\sigma} \mapsto \vec{\sigma} - 2 \pi k \vec{\nu}_N$ spanning a surface whose boundary is the loop $C$.
Every term in (\ref{eq:tH_loop_dualphotonEFT_lifted}) requires such a nontrivial kink, because the root-charge Wilson loop cannot compensate for the shift ($\vec{\sigma} \mapsto \vec{\sigma} - 2 \pi k \vec{\nu}_N$) due to the nontrivial $N$-ality.
Therefore, we obtain
\begin{align}
    \braket{H^{(\vec{\sigma})}(C; \Sigma)}_{\vec{\sigma} = \vec{\sigma}_k} =  (\text{area-law terms}),~~~\text{for}~k\neq0.
\end{align}
Incidentally, the string tension of $H^{(\vec{\sigma})}(C; \Sigma)$, or the dual string tension, is equal to the smallest string tension of the Wilson loop of $N$-ality $k$, which is the tension of the Wilson loop $W_{\vec{\mu}_k} (C)$\footnote{The Wilson loop of the $k$-index antisymmetric representation has the minimal string tension in the semiclassical description.
For details of the string tensions of the Wilson loop in the deformed Yang-Mills theory, see \cite{Poppitz:2017ivi}.}.

We then extend our consideration to the low-energy behaviors of a general dyonic line: $\braket{H^{(\vec{\sigma})}(C; \Sigma) W_{\vec{\mu}} (C)}$.
Because all root vectors are summed in (\ref{eq:tH_loop_dualphotonEFT_lifted}), only the $N$-ality of $\vec{\mu}$ is relevant, and we write the $N$-ality of $\vec{\mu}$ by $[\vec{\mu}]_{\mathbb{Z}_N}$. 
As the Wilson loop is the monodromy defect for $\vec{\sigma}$, we can rephrase $W_{\vec{\mu}} (C)$ as the change of the lift $\vec{\tilde{\sigma}} \mapsto \vec{\tilde{\sigma}} + 2\pi \vec{\mu}$ on a surface $\Sigma$.
Thus, the Wilson loop $W_{k\vec{\nu}_N} (C)$ compensates for the permutation $P_W^{-1} \vec{\tilde{\sigma}}_k = \vec{\tilde{\sigma}}_k - 2 \pi k \vec{\nu}_N$ in this lift $\vec{\tilde{\sigma}} = \vec{\sigma}_k$, implying a perimeter law for $\braket{\tilde{U}_{P_W^{-1}}^{(\vec{\sigma})}(\Sigma) W_{k\vec{\nu}_N} (C)}_{\vec{\tilde{\sigma}}= \vec{\sigma}_k}$.
To sum up, we obtain,
\begin{align}
    \braket{H^{(\vec{\sigma})}(C; \Sigma) W_{\vec{\mu}}(C)}_{\vec{\sigma} = \vec{\sigma}_k} = \begin{cases}
        1+(\text{area-law terms})~~~&\text{for}~[\vec{\mu}]_{\mathbb{Z}_N} = k \\
        (\text{area-law terms})~~~&\text{for}~[\vec{\mu}]_{\mathbb{Z}_N} \neq k 
    \end{cases} \label{eq:dyonic_loop_k-th_vacuum}
\end{align}
The appearance of the perimeter law can also be regarded as the phenomenon in which one of the kinks emitted by $H^{(\vec{\sigma})}(C; \Sigma) $ can be absorbed by the Wilson loop $W_{\vec{\mu}}(C)$ if the $N$-ality is matched. 

This result is consistent with the Wilson-'t Hooft classification for the dyonic lines: 
When $|\theta - 2\pi k| < \pi$, the perimeter-law dyonic operator is $H^{(\vec{\sigma})}(C; \Sigma) W_{\vec{\mu}}(C)$ with $N$-ality $[\vec{\mu}]_{\mathbb{Z}_N} = k $. 
This implies that the confining vacua for $|\theta - 2\pi k| < \pi$ belong to the level-$k$ SPT state of the $\mathbb{Z}_N^{[1]}$ symmetry. 

\subsection{Temporal loops}

For completeness, we also examine the temporal loops.
The temporal Wilson loop is the Polyakov loop, and the holonomy potential stabilizes the $\left( \mathbb{Z}_N^{[0]} \right)_{\mathrm{3d}}$ center symmetry: $\operatorname{tr}(P_4) = 0$ in our setup.

We have observed that the non-genuine temporal 't Hooft loop $H(\{p,p'\};\gamma)$ is defined by (\ref{eq:def_temp_tHloop}) in the 3d effective theory.
We immediately see that 
\begin{align}
     \begin{cases}
        \braket{H(\{p,p'\};\gamma)}_{\vec{\sigma} = \vec{\sigma}_k}  \neq 0~~~&\text{for}~k=0, \\
        \braket{H(\{p,p'\};\gamma)}_{\vec{\sigma} = \vec{\sigma}_k}  =0~~~&\text{for}~k\neq 0 .
    \end{cases} 
\end{align}

Next, let us consider the dyonic loops.
In the 3d monopole semiclassics, where $\operatorname{tr}(P_4) = 0$, a good order parameter cannot be constructed simply by attaching $\operatorname{tr}(P_4)$ to $H(\{p,p'\};\gamma)$.
We need another dyonic line operator, which has proper $\left( \mathbb{Z}_N^{[0]} \right)_{\mathrm{3d}}$ charge and serves as the boundary of a $\left( \mathbb{Z}_N^{[1]} \right)_{\mathrm{3d}}$ defect.
Such an operator can be constructed by replacing $\vec{\sigma} \mapsto \vec{\sigma} + k' \vec{\phi}$ in (\ref{eq:def_temp_tHloop})\footnote{In constructing such a dyonic operator, it is crucial to properly assign charge-$k'$ of the $\left( \mathbb{Z}_N^{[0]} \right)_{\mathrm{3d}}$ symmetry at $\{ p, p' \}$.
On this point, averaging the permutation at $\{ p, p' \}$ independently is essential; see remark below (\ref{eq:def_temp_tHloop}). 
}:
\begin{align}
    \left( HW^{k'} \right)(\{p,p'\};\gamma) := \sum_{j,j'=1}^{N}  \rme^{\im  \vec{\alpha}_{j',j} \cdot (\vec{\sigma}(p) + k' \vec{\phi}(p))} \rme^{\im \int_\gamma \vec{\nu}_j \cdot d (\vec{\sigma} + k' \vec{\phi})},
\end{align}
which has a charge-$k'$ of the $\left( \mathbb{Z}_N^{[0]} \right)_{\mathrm{3d}}$ symmetry at $\{ p, p' \}$.
By construction, we obtain 
\begin{align}
     \begin{cases}
        \braket{\left( HW^{k'} \right)(\{p,p'\};\gamma)}_{\vec{\sigma} = \vec{\sigma}_k}  \neq 0~~~&\text{for}~k=k', \\
        \braket{\left( HW^{k'} \right)(\{p,p'\};\gamma)}_{\vec{\sigma} = \vec{\sigma}_k}  =0~~~&\text{for}~k\neq k' ,
    \end{cases} 
\end{align}
which is consistent with the Wilson-'t Hooft classification. 
Importantly, the dressed electric charge for the temporal dyonic loop with the long-range order is identical to the one for the spatial dyonic loop with the perimeter law for each confinement vacuum.

\section{Dual string tension in the thermal deconfined phase}
\label{sec:tHooft_thermal_deconf}

In the thermal deconfined phase, there exist $N$ vacua associated with the spontaneous $\left( \mathbb{Z}_N^{[0]} \right)_{\mathrm{3d}}$ symmetry breaking, and we can consider the $\mathbb{Z}_N^{[0]}$ domain wall connecting these vacua. 
In this phase, the spatial dyonic loops are confined since deconfinement on the electric side along the temporal direction is dual to the confinement on the magnetic side on the spatial directions. 
We first clarify the precise connection between the $\mathbb{Z}_N$ domain wall and the dual confining strings for the spatial dyonic loops: There exists a finer classification compared with the one discussed in previous literature. 

We then confirm the consequence of the finer classification using the semiclassical computation. As we explain below, the 3d monopole semiclassics for the mass-deformed $\mathcal{N}=1$ supersymmetric Yang-Mills (SYM) theory has the confinement-deconfinement phase transition caused by the fermion mass parameter, and this phase transition is expected to be smoothly connected to the thermal confinement-deconfinement phase transition. 
We employ the definition of the screened 't~Hooft loop in the $\mathbb{Z}_N^{[0]}$-broken phase and explicitly compute dual string tensions.

\subsection{Refined classification of \texorpdfstring{$\mathbb{Z}_N$}{Z(N)} domain wall}

In the $0$-form center-broken phase, the spatial 't Hooft loop $H^{(\vec{\sigma})}(C; \Sigma)$ relates different $\left( \mathbb{Z}_N^{[0]} \right)_{\mathrm{3d}}$-broken vacua by its definition, and thus it generates the  $\left( \mathbb{Z}_N^{[0]} \right)_{\mathrm{3d}}$ domain wall.
This equivalence of the dual string tension and the tension of the $\left( \mathbb{Z}_N^{[0]} \right)_{\mathrm{3d}}$ domain wall has been discussed in \cite{Korthals-Altes:1999cqo}. 
While this argument appears reasonable, we should note that it is not precise; the open $\left( \mathbb{Z}_N^{[0]} \right)_{\mathrm{3d}}$ defects are classified into $N$ types, $H^{(\vec{\sigma})}(C; \Sigma) W_{\vec{\mu}}(C)$ for $[\vec{\mu}]_{\mathbb{Z}_N} = 0,1,\cdots,N-1$.
Hence, the true $\left( \mathbb{Z}_N^{[0]} \right)_{\mathrm{3d}}$ domain wall tension $T_{\mathrm{DW}}$ is determined by
\begin{align}
    T_{\mathrm{DW}} = \operatorname{min}_{\vec{\mu}} \left( T_{H^{(\vec{\sigma})}(C; \Sigma) W_{\vec{\mu}}(C)} \right)= \operatorname{min}_{[\vec{\mu}]_{\mathbb{Z}_N} \in \mathbb{Z}_N} \left( T_{H^{(\vec{\sigma})}(C; \Sigma) W_{\vec{\mu}}(C)} \right) ,
\end{align}
where $T_{H^{(\vec{\sigma})}(C; \Sigma) W_{\vec{\mu}}(C)}$ denotes the string tension of $H^{(\vec{\sigma})}(C; \Sigma) W_{\vec{\mu}}(C)$.

Let us first give a refined classification of the $(\mathbb{Z}_N^{[0]})_{\mathrm{3d}}$ domain wall from the viewpoint of the symmetry in this section. In the next subsection, we confirm the prediction based on the symmetry using the $3$d semiclassics for the mass-deformed $\mathcal{N}=1$ SYM theory. 
In particular, we will see that the domain-wall state encounters a phase transition as $\theta$ is shifted by $2\pi$ while the bulk phase changes smoothly. 

The domain-wall tension $T_{\mathrm{DW}}$ can be extracted from the volume dependence of the partition function with the twisted boundary condition for the broken symmetry~\cite{Maeda:2025ycr}:
\begin{equation}
    Z_{\mathrm{twisted}} \sim \exp\left( - T_{\mathrm{DW}} \, |\mathrm{Area}| \right).
\end{equation}
Here, the theory is put on a compactified space, e.g., $T^3 \times S^1_L$, and the size of the $3$-torus $T^3$ is supposed to be much larger than the size of the thermal circle, $S^1_L$. 
We impose a $\left( \mathbb{Z}_N^{[0]} \right)_{\mathrm{3d}}$ twisted boundary condition along a  $1$-cycle of the spatial torus $T^3$ and take the large-volume limit. The factor $|\mathrm{Area}|$ denotes the area of the two spatial directions orthogonal to the twisted direction.
Let us denote the direction with the $\left( \mathbb{Z}_N^{[0]} \right)_{\mathrm{3d}}$ twist as $x_3$. 
From the 4d perspective, this corresponds to introducing a 1-form symmetry flux along the $x_3$-$x_4$ plane: $\int_{S^1_{(3)}}A_{3\diff}=\int_{T_{(34)}^2} B_{\mathrm{4d}} = \frac{2 \pi}{N}$, where $T_{(34)}^2$ is the $x_3$-$x_4$ torus\footnote{In this sense, the $\left( \mathbb{Z}_N^{[0]} \right)_{\mathrm{3d}}$ domain wall might be understood as a ``center vortex'' in the 4d viewpoint.
This would be consistent with the observation that the 't Hooft loop would be a boundary of center vortex \cite{Reinhardt:2002mb}.
}.
The domain wall is formed along the $x_1$-$x_2$ torus, $T_{(12)}^2$.

We note that the $(\mathbb{Z}_N^{[0]})_{\mathrm{3d}}$ domain wall is a $2$d extended object. 
If there exists an unbroken $1$-form symmetry in the $3$d effective theory, such a domain-wall state may acquire the local counterterm for the background gauge field $B_{3\diff}$ for this $1$-form symmetry. 
Then, the possible behaviors of the partition function with its background gauge field $B_{3\diff}$ should be given by 
\begin{equation}
    Z_{\mathrm{twisted}} \sim \exp\left( - T_{\mathrm{DW}}^{(k)}(\theta) \, |\mathrm{Area}| + \im k \int_{T^2_{(12)}} B_{\mathrm{3d}} \right),
    \label{eq:DomainWall_Z}
\end{equation}
with some $k=0,1,\ldots, N-1$ (mod $N$). 
When the domain wall carries the level-$k$ $2$d SPT phase in this manner, creating a boundary of the domain wall requires a line operator with a 1-form charge $k$ at the boundary. Therefore, the domain wall tension must coincide with the tension of the Wilson-'t Hooft loop $H(C; \Sigma) W^{-k}(C)$\footnote{Here, we suppress superscript/subscript as the kinematical discussion here is not limited to the 3d monopole semiclassics.}. 
We now obtain the refined classification of the domain wall: The $\left( \mathbb{Z}_N^{[0]} \right)_{\mathrm{3d}}$ domain walls are classified by the 2d SPT phase $\exp (\im k \int_{T^2_{(12)}} B_{\mathrm{3d}})$, and this domain wall corresponds to the Wilson-'t Hooft loop $H(C; \Sigma) W^{-k}(C)$.

In the context of the twisted partition function, the mixed anomaly yields a non-trivial prediction for the domain walls. 
Recall the mixed anomaly between the $4$d 1-form symmetry and the $2\pi$-periodicity of the $\theta$-term gives the following relation for the $3$d partition function,
\begin{equation}
    Z_{\theta+2\pi}[A_{\mathrm{3d}},B_{\mathrm{3d}}] = \exp\left( \frac{\im N}{2\pi}\int A_{\mathrm{3d}} \wedge B_{\mathrm{3d}} \right) Z_{\theta}[A_{\mathrm{3d}},B_{\mathrm{3d}}].
    \label{eq:3dAnomaly}
\end{equation}
Let us assume that the domain-wall states at $\theta+2\pi$ and at $\theta$ has the SPT levels, $k$ and $k'$, respectively, and then substituting the formula~\eqref{eq:DomainWall_Z} into the anomaly relation~\eqref{eq:3dAnomaly} gives 
\begin{align}
    &\exp\left( - T_{\mathrm{DW}}^{(k)}(\theta+2\pi) \, |\mathrm{Area}| + \im k \int_{T^2_{(12)}} B_{\mathrm{3d}} \right) \notag\\
    &= \exp\left( - T_{\mathrm{DW}}^{(k')}(\theta) \, |\mathrm{Area}| + \im (k'+1) \int_{T^2_{(12)}} B_{\mathrm{3d}} \right). 
\end{align}
Comparison of the phase factors implies $k'=k-1$, and thus the mixed-anomaly equation dictates the following relation for the domain wall tension:
\begin{equation}
    T_{\mathrm{DW}}^{(k)}(\theta+2\pi) = T_{\mathrm{DW}}^{(k-1)}(\theta) \label{eq:anomaly_relation_DW}
\end{equation}
Thus, the domain-wall state must encounter a phase transition at least once when we gradually increase the $\theta$ angle by $2\pi$. 
In terms of Wilson-'t Hooft loops, this is precisely the Witten effect: $H(C; \Sigma) \xrightarrow[]{~\theta \mapsto \theta + 2\pi~}H(C; \Sigma) W(C)$.

The realization of anomaly on the domain wall would be an interesting subject.
For example, the dynamics of the domain wall for $\mathcal{N}=1$ SYM and massless QCD(adj) was studied in Ref.~\cite{Anber:2018xek}.

\subsection{\texorpdfstring{$\mathcal{N}=1~SU(2)$}{N=1 SU(2)} super Yang-Mills theory with mass deformation}

To illustrate the above refined properties for the domain-wall states, we consider the $\mathcal{N}=1$ super Yang-Mills theory with the fermion mass deformation, which realizes the theoretically controlled confinement-deconfinement bulk phase transition. 
This confinement-deconfinement phase transition is believed to be continuously connected to the thermal phase transition in the pure Yang-Mills theory (see \cite{Poppitz:2012sw, Poppitz:2012nz, Anber:2014lba}).
To avoid technical complications, let us focus on the $\mathcal{N}=1$ $SU(2)$ super Yang-Mills theory with the fermion mass deformation $m$.
We parameterize the dual photon and holonomy $(\sigma,\varphi)$ as
\begin{align}
    \vec{\phi} = \vec{\phi}_c + \frac{g^2}{4 \pi} \varphi \vec{\mu}_1,~~~\vec{\sigma} = \sigma \vec{\mu}_1,
\end{align}
with periodicity $\sigma \sim \sigma + 2\pi$.
Note that the root vector is $\vec{\alpha}_1 = 2 \vec{\mu}_1$, so the root periodicity is twice that of the weight lattice.

The 3d monopole and bion potential of $(\sigma,\phi)$ is\footnote{We only use this effective theory within $|\vec{\phi} -\vec{\phi}_c| \lesssim O(g^2)$, hence, the GPY potential becomes a higher order correction.} \cite{Poppitz:2012sw, Anber:2014lba}, 
\begin{align}
    V_{\mathrm{monopole/bion}} (\sigma,\varphi) =& 4V_0 \cosh(2 \varphi) - 4 V_0 \cos (2 \sigma) \notag \\
    &~~~ - \gamma V_0 \left[ \rme^{-\varphi}\cos (\sigma + \theta/2) + \rme^{\varphi}\cos (\sigma - \theta/2) \right] \label{eq:simpleSYMpotential}
\end{align}
where
\begin{align}
    V_0 = \frac{9N^2}{16\pi^2}\frac{L^3 \Lambda^6}{g^2},~~ \gamma=\frac{32\pi^2}{3N^2} \frac{m}{L^2 \Lambda^3}.
\end{align}
The first two terms represent the bions, molecules of monopole and antimonopole, and the second term is the monopole contribution.
At a certain point\footnote{This point can be found by testing the stability of the confining vacuum $(\sigma,\varphi)  =(0,0),~(\pi,0)$ since the phase transition for the case of $SU(2)$ is of the $2$nd order. The confinement-deconfinement transition point turns out to be independent of the $\theta$ parameter within our approximation, while its dependence comes in when we include the one-loop GPY potential~\cite{Chen:2020syd}. } $\gamma = 8$, the deconfinement transition happens.
Thus, the mass deformation for the adjoint fermion plays the role of the ``temperature'': For $\gamma<8$, the system is in the confined phase, and the vacuum is described by 
\begin{align}
        (\sigma,\varphi)  =
    \begin{cases}
        (0,0)&\text{for}~-\pi< \theta < \pi, \\
        (\pi,0)&\text{for}~\pi < \theta < 3\pi. 
    \end{cases}
\end{align}
For $\gamma>8$, the system is in the deconfined phase, and let us write the $(\mathbb{Z}_2^{[0]})_{\mathrm{3d}}$-broken vacua as 
\begin{align}
    (\sigma,\varphi)  =
        (\sigma_*(\gamma,\theta),\varphi_*(\gamma,\theta)), ~ (-\sigma_*(\gamma,\theta),-\varphi_*(\gamma,\theta)). 
\end{align}
The confinement-deconfinement transition at $\gamma=8$ is given by the $2$nd-order phase transition for the case of $SU(2)$.

Now, let us observe the spatial loop operators in these phases.
In the confining phases, $(\sigma,\varphi)  =(0,0),~(\pi,0)$, the previous section shows that the 't Hooft loop and the Wilson-'t Hooft loop behave as
\begin{align}
    \braket{H^{(\vec{\sigma})}(C; \Sigma) } = \begin{cases}
        1+(\text{area-law terms})~~~&\text{for}~0\leq \theta < \pi \\
        (\text{area-law terms})~~~&\text{for}~~\pi < \theta < 2\pi 
    \end{cases} \\
   \braket{H^{(\vec{\sigma})}(C; \Sigma) W(C) } = \begin{cases}
        (\text{area-law terms})~~~&\text{for}~0\leq \theta < \pi \\
        1+(\text{area-law terms})~~~&\text{for}~~\pi < \theta < 2\pi 
    \end{cases} 
\end{align}

Let us focus on the deconfined phase $(\sigma,\varphi)  = \pm (\sigma_*(\gamma,\theta),\varphi_*(\gamma,\theta))$ in the following.
In principle, the 't Hooft loop $\left(H(C; \Sigma) \right)_{\mathrm{screened}} $ contains many terms, but the same argument above (\ref{eq:tH_loop_dualphotonEFT}) applies as long as $\vec{\phi} \approx\vec{\phi}_c$.
From a parallel argument (by ignoring the terms passing singularities $\vec{\phi} =0,\pi \vec{\alpha}_1$), in a lift $(\tilde{\sigma},\tilde{\varphi}) \in  \mathbb{R}\times \mathbb{R}$, we can express the screened 't Hooft loop as
\begin{align}
    \left(H(C; \Sigma) \right)_{\mathrm{screened}} \simeq \sum_{n \in \mathbb{Z}}\tilde{U}_{\tilde{\sigma} \mapsto -\tilde{\sigma}}(\Sigma) \tilde{U}_{\tilde{\varphi} \mapsto -\tilde{\varphi}}(\Sigma) W^{2n} (C) .
\end{align}
where $\tilde{U}_{\tilde{\sigma} \mapsto -\tilde{\sigma}}(\Sigma)$ is the defect flipping $\tilde{\sigma} \mapsto -\tilde{\sigma}$ on the open surface $\Sigma$, and $\tilde{U}_{\tilde{\varphi} \mapsto -\tilde{\varphi}}(\Sigma)$ is that of $\tilde{\varphi} \mapsto -\tilde{\varphi}$ on $\Sigma$.
The presence of a term $\tilde{U}_{\tilde{\sigma} \mapsto -\tilde{\sigma}}(\Sigma) \tilde{U}_{\varphi \mapsto -\varphi}(\Sigma) W^{2n} (C)$ necessitates a kink from $(\sigma,\varphi)  = (\pm \sigma_*,\pm \varphi_*)$ to $(\sigma,\varphi)  = (\mp\sigma_*+ 4\pi n,\mp \varphi_*)$.
By choosing $-\pi < \sigma_* < + \pi$, the dominant term is a kink from $(\sigma,\varphi)  = \pm (\sigma_*,\varphi_*)$ to $(\sigma,\varphi)  = \mp( \sigma_*, \varphi_*)$.
Therefore, the spatial 't Hooft loop shows the area law, and the dual string tension is,
\begin{align}
    \braket{\left(H(C; \Sigma) \right)_{\mathrm{screened}}} \sim \rme^{-  T_{(\sigma_*,\varphi_*) \mapsto -( \sigma_*, \varphi_*)}  \operatorname{Area}(C)},
\end{align}
where $T_{(\sigma_*,\varphi_*) \mapsto -( \sigma_*, \varphi_*)}$ denotes the tension of the kink $(\sigma_*,\varphi_*) \mapsto -( \sigma_*, \varphi_*)$ and $\operatorname{Area}(C)$ means the minimal area whose boundary is $C$.
This tension is the so-called dual string tension: $T_{H} = T_{(\sigma_*,\varphi_*) \mapsto -( \sigma_*, \varphi_*)}$.

\begin{figure}[t]
\centering
\begin{minipage}{.32\textwidth}
  \subfloat[$\theta = 0$]{
    \includegraphics[width=\textwidth]{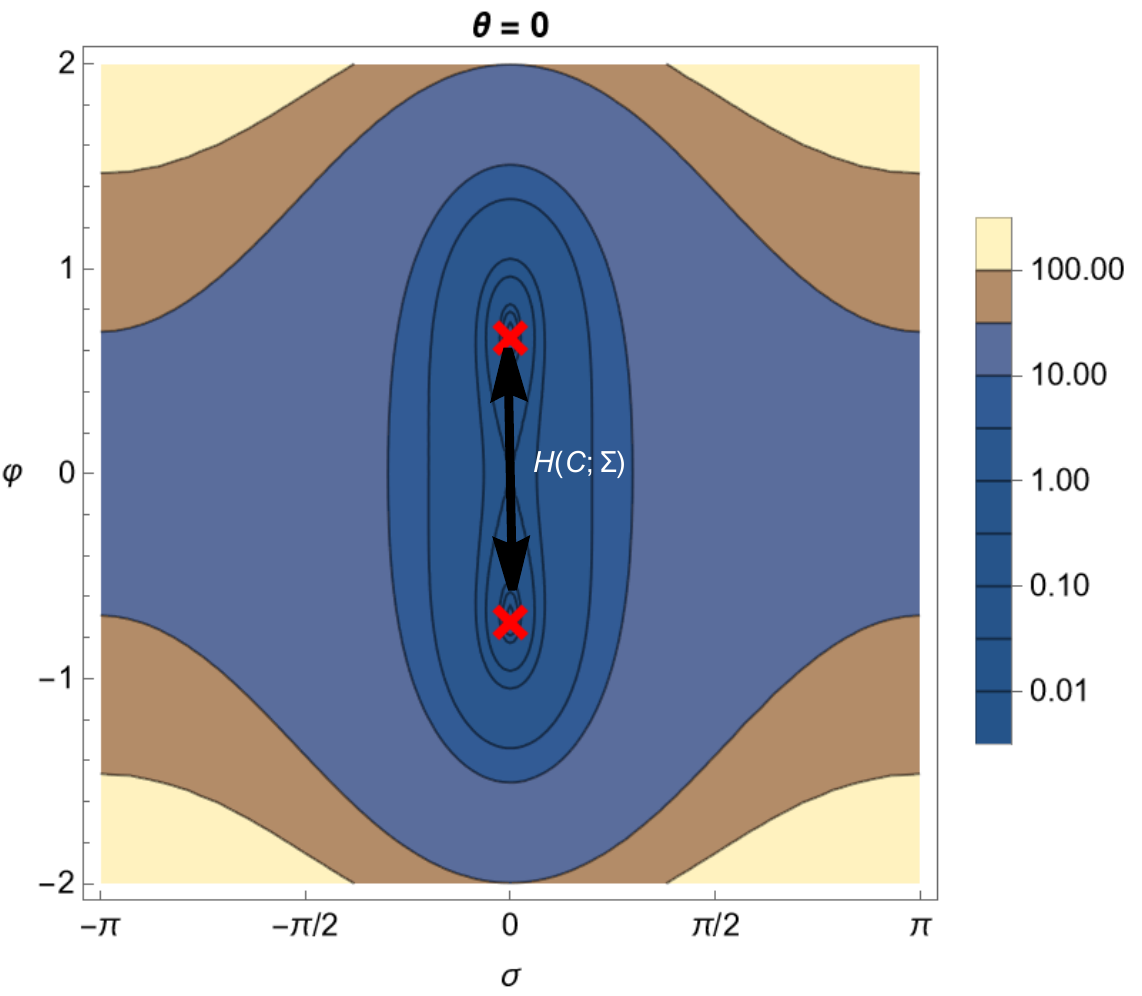}
  }
\end{minipage}\quad
\begin{minipage}{.32\textwidth}
  \subfloat[$\theta = \pi/2$]{
    \includegraphics[width=\textwidth]{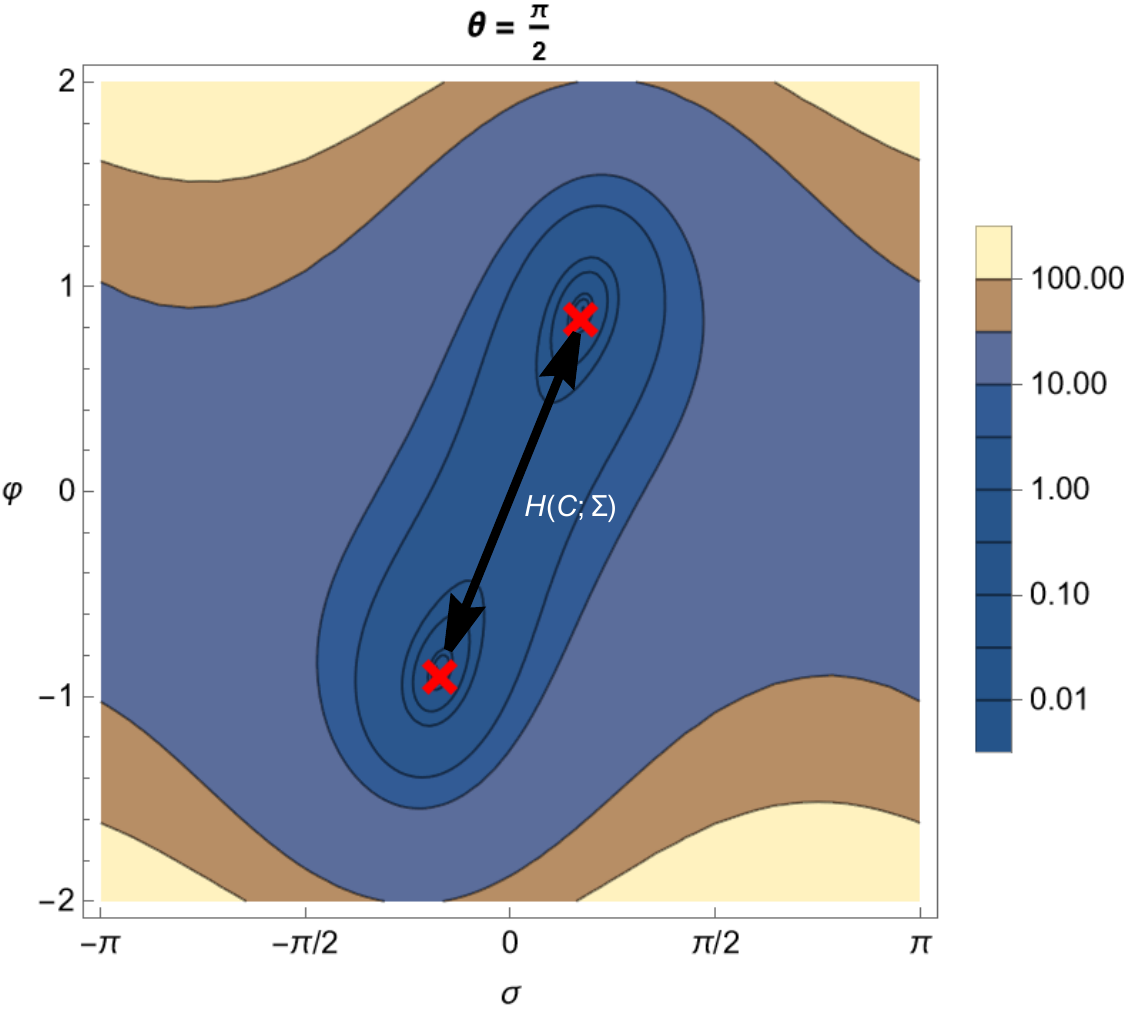}
  } 
\end{minipage}
\begin{minipage}{.32\textwidth}
  \subfloat[$\theta = 3 \pi/2$]{
    \includegraphics[width=\textwidth]{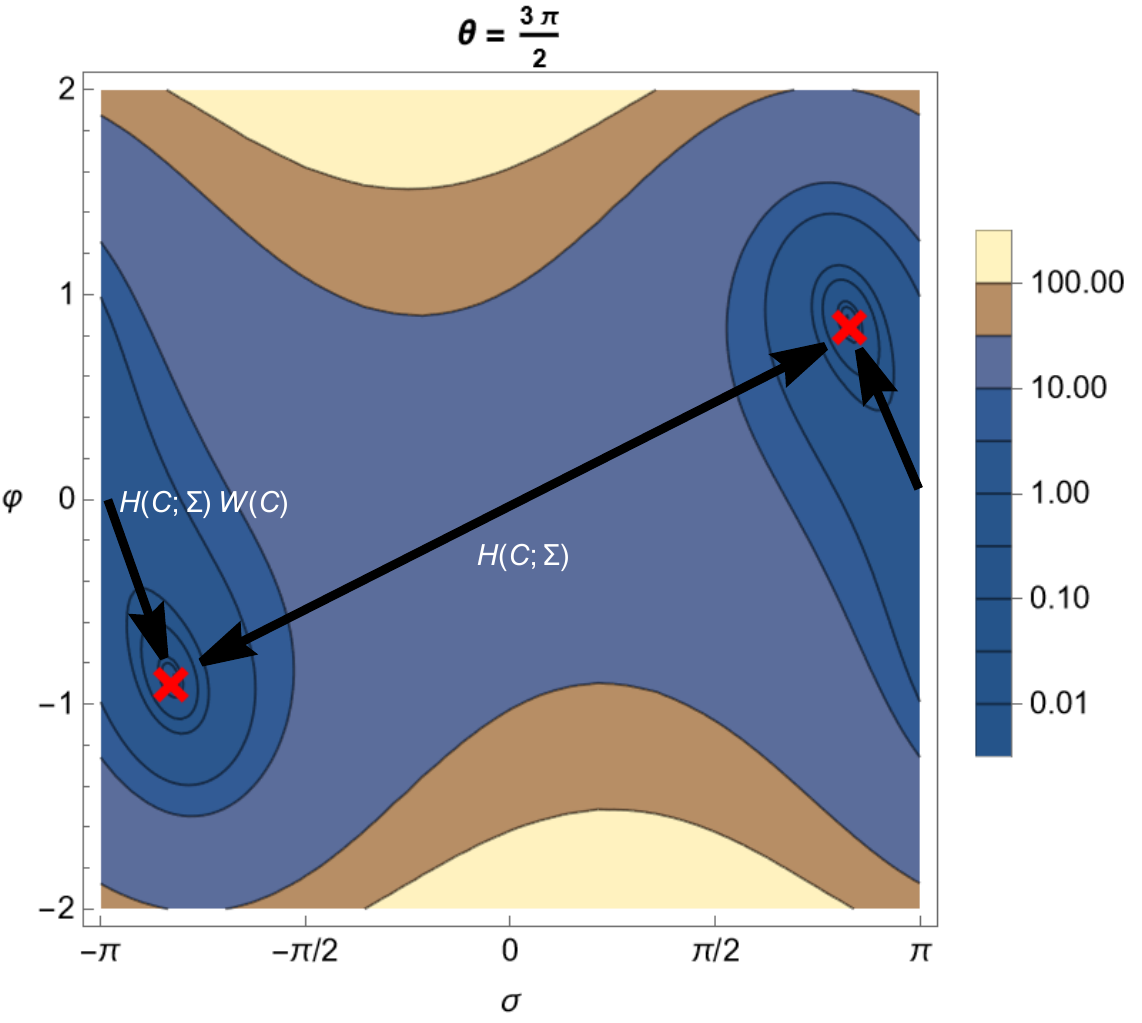}
  } 
\end{minipage}
\caption{Contour plots of the potential $V(\varphi, \sigma)/V_0$ with $\gamma = 10$ for representative values of $\theta$: $\theta = 0, ~ \pi/2$, and $3 \pi/2$. The minima are indicated by cross symbols. 
For visual clarity, the potential minimum is set to zero in these plots.
At this mass deformation parameter $\gamma = 10$, the $\mathbb{Z}_2$ center symmetry is spontaneously broken. 
Kink configurations interpolating between two vacua correspond to domain walls. The insertion of a spatial 't Hooft loop $H(C;\Sigma)$ induces a kink on the surface $\Sigma$ connecting $(\sigma, \varphi) = (\sigma_*, \varphi_*)$ and $(-\sigma_*, -\varphi_*)$. For $(-\pi<)\theta < \pi$, this configuration represents the kink with the minimum tension. However, as illustrated in the plot for $\theta = 3 \pi/2$, in the range $\pi < \theta (< 3\pi)$, a different kink connecting $(\sigma, \varphi) = (\sigma_*, \varphi_*)$ and $(2\pi-\sigma_*, -\varphi_*)$ has the minimum tension. Since the $2\pi$ shift of $\sigma$ corresponds to attaching a Wilson loop to the boundary, the lightest kink in this regime is generated by the composite operator $H(C;\Sigma) W(C)$. }
\label{fig:deconf_potential}
\end{figure}

In Fig.~\ref{fig:deconf_potential}, we show the contour plot of the effective potential (\ref{eq:simpleSYMpotential}) for the deconfined phase, $\gamma=10>8$, in the $\sigma$-$\varphi$ plane. 
As we can see from the figure, the minimum $(\sigma_*,\varphi_*)$ of $\varphi_* > 0$ is located at 
\begin{align}
    \begin{cases}
        0<\sigma_*<\pi/2~~\mathrm{for}~0<\theta<\pi, \\
        \pi/2<\sigma_*<\pi~~\mathrm{for}~\pi<\theta< 2 \pi. 
    \end{cases}
\end{align}
For $0<\theta<\pi$, the minimal $\left(\mathbb{Z}_N^{[0]}\right)_{\mathrm{3d}}$ domain wall connects the vacua $(\sigma_*,\varphi_*)$ and $(-\sigma_*,-\varphi_*)$, which is nothing but the kink configuration appearing in $(H(C;\Sigma))_{\mathrm{screened}}$. 
However, for $\pi<\theta< 2 \pi$, the point $(\sigma,\varphi)  =  (\sigma_*,\varphi_*)$ is closer to $(\sigma,\varphi)  =  (2 \pi -\sigma_*,-\varphi_*)$ rather than $(\sigma,\varphi)  =  ( -\sigma_*,-\varphi_*)$.
In this region $\pi<\theta< 2 \pi$, the minimal $\left( \mathbb{Z}_N^{[0]} \right)_{\mathrm{3d}}$ domain wall is generated by $\left(H(C; \Sigma) W(C) \right)_{\mathrm{screened}} $, of which the tension is given by the corresponding kink: $T_{HW} = T_{(\sigma_*,\varphi_*) \mapsto (2\pi - \sigma_*, -\varphi_*)}$.
Thus, the domain-wall state has the phase transition at $\theta=\pi$, while the bulk state is smooth in $\theta$ for the deconfined phase.\footnote{By adding $O(g^2)$ corrections, the $CP$-broken deconfined phase appears in a certain window of $\gamma$, as pointed out in \cite{Chen:2020syd}, and the discussion here should be modified when $\gamma$ is inside this finite window.
In this case, rather than the level crossing of the string tensions $T_{H}$ and $T_{HW}$, a sudden jump of the wall tension occurs at $\theta = \pi$ associated with the bulk phase transition: $\left( T_{H}\right)_{\theta = \pi -0}=\left( T_{HW}\right)_{\theta = \pi +0}$ but $\left( T_{H}\right)_{\theta = \pi -0} \neq \left( T_{HW}\right)_{\theta = \pi -0}$.
} 

Our observation gives a concrete realization of the anomaly relation for the domain wall tension (\ref{eq:anomaly_relation_DW}):
\begin{align}
    T_{H}(\theta + 2\pi) = T_{HW}(\theta),~~~T_{HW}(\theta + 2\pi) = T_{H}(\theta),
\end{align}
Here, $T_{H}(\theta)$ corresponds to $T_{\mathrm{DW}}^{(k=0)}(\theta) $ and $T_{HW}(\theta)$ corresponds to $T_{\mathrm{DW}}^{(k=1)}(\theta)$ in the notation of (\ref{eq:anomaly_relation_DW}).
The lightest $\left( \mathbb{Z}_N^{[0]} \right)_{\mathrm{3d}}$ domain wall tension $T_{\mathrm{DW}}$ is,
\begin{align}
    T_{\mathrm{DW}}(\theta) = \operatorname{min} (T_H(\theta), T_{HW}(\theta))=
    \begin{cases}
        T_{H}(\theta)  ~~\mathrm{for}~0<\theta<\pi, \\
        T_{HW}(\theta)  ~~\mathrm{for}~\pi<\theta< 2 \pi,
    \end{cases}
    \label{eq:DW_level_crossing}
\end{align}
and thus the two domain-wall tensions exhibit the level-crossing phenomenon and are degenerate at $\theta=\pi$, which is a consequence of the $CP$ symmetry.

\section{Summary and outlook}
\label{sec:Summary}

In this paper, we have revisited the definition of the 't~Hooft loop operator in the context of the $3$d monopole semiclassics on $\mathbb{R}^3\times S^1$. 
Since the $3$d monopole semiclassics is based on the adjoint Higgsing, $SU(N)\xrightarrow{\mathrm{Higgs}}U(1)^{N-1}$, there exists a standard definition of the 't~Hooft loop operator, \eqref{eq:spatial_tHt_def}, which has been also used in previous literature. 
We first point out that its expectation value cannot be computed in a well-defined way within the $3$d monopole semiclassics $\mathcal{L}_{\mathrm{3d}}^{(\vec{\sigma},\vec{\phi})}$, \eqref{eq:3dEFT_sigmaphi}, since it imposes the twisted boundary condition that necessarily requires the singular point, where the adjoint Higgsing fails. 
If we forcibly proceed with the computation neglecting the singularity, we obtain the area law for all the dyonic loop operators, which contradicts the expected behavior of the Wilson-'t~Hooft classification for the confinement phases. 

To resolve this puzzle, we introduce the twist vortex operator, $T_{\mathrm{twist}}(C)$, in \eqref{eq:twisted_vortex_expression}, to consider the screening phenomena of the 't~Hooft loop by the twist vortices. 
Since the $3$d low-energy effective theory has accidentally enhanced $1$-form symmetries, there exists the strong selection rule for the loop operators, which is not present in the full UV description, and we need to take into account the screening phenomenon of the loop operators explicitly to obtain the correct low-energy behaviors. 
This observation leads to the screened 't~Hooft loop, $(H(C;\Sigma))_{\mathrm{screened}}$, and we obtain the new definition~\eqref{eq:tH_loop_dualphotonEFT} of the 't~Hooft loop, $H^{(\vec{\sigma})}(C;\Sigma)$, for the monopole semiclassics. 
Most importantly, this operator accepts the well-defined computation within the $3$d Abelianized effective theory unlike the case of the naive definition~\eqref{eq:spatial_tHt_def}. 

Using the new definition of the 't~Hooft loop, we compute the expectation values of the dyonic loop operators, $H^{(\vec{\sigma})}W_{\vec{\mu}}$, for each confinement vacuum at $|\theta-2\pi k|<\pi$. 
We then find in \eqref{eq:dyonic_loop_k-th_vacuum} that it shows the perimeter law only if $[\vec{\mu}]_{\mathbb{Z}_N}=k$ and the other cases show the area law.
This is exactly what is predicted by the Wilson-'t~Hooft classification, and the level $k$ of the SPT state for the unbroken $(\mathbb{Z}_N^{[1]})_{\mathrm{4d}}$ symmetry is given by the electric charge of the deconfined dyonic (non-genuine) loop. 

We then extend our discussion to the case of the thermal deconfined phase and give the refined classification of the $(\mathbb{Z}_N^{[0]})_{\mathrm{3d}}$ domain wall. 
In the deconfined phase, the bulk state is smooth as we increase the $\theta$ angle by $2\pi$, but the anomaly relation tells the domain-wall state should encounter the phase transition, \eqref{eq:anomaly_relation_DW}, as it acquires the $2$d local counterterm, $\im \int_{\mathrm{wall}}B_{\mathrm{3d}}$, for the unbroken $3$d $1$-form symmetry. 
To illustrate this phenomenon, we consider the mass-deformed $\mathcal{N}=1$ SYM theory on $\mathbb{R}^3\times S^1$, where the fermion mass causes the confinement-deconfinement transition and plays the role of the temperature for the thermal Yang-Mills case. 
We explicitly compute the dual string tensions for $H^{(\vec{\sigma})}(C;\Sigma)$ and $H^{\vec{(\sigma)}}(C;\Sigma)W(C)$ and confirm the prediction of the symmetry and anomaly.

Lastly, let us present several future prospects that can come out of this study:

\begin{itemize}
    \item  The $\mathcal{N}=1^*$ SYM (the $\mathcal{N}=4$ SYM softly-broken to a $\mathcal{N}=1$ theory) offers a quite interesting theoretical playground that realizes all possible gapped ground states with the $\mathbb{Z}_N^{[1]}$ symmetry~\cite{Donagi:1995cf}.
    The 3d description of the $\mathcal{N}=1^*$ SYM on $\mathbb{R}^3\times S^1$ has been developed in \cite{Dorey:1999sj}, and its superpotential is given by the Weierstrass elliptic function. 
    To our knowledge, it has not been explored if the low-energy behaviors of the loop operators are consistent with the expected ones from the Wilson-'t~Hooft classification, and we argue that twist vortices play a pivotal role to obtain the correct behaviors there. 

An immediate application of the twist vortex is the perimeter law of the spatial Wilson loop in the Higgs phase of the $\mathcal{N}=1^*$ $SU(2)$ SYM.
According to Dorey's 3d description \cite{Dorey:1999sj}, the Higgs phase is described by $(\sigma, \phi) = (\pi, 0)$.
In the Higgs phase, we suppose that $\left( \mathbb{Z}_N^{[1]} \right)_{\mathrm{4d}}$ is spontaneously broken.
Whereas the spontaneous breakdown of the temporal part $\left( \mathbb{Z}_N^{[0]} \right)_{\mathrm{3d}}$ is clear, that of the spatial part $\left( \mathbb{Z}_N^{[1]} \right)_{\mathrm{3d}}$ is nontrivial because 
the spatial Wilson loop is typically defined as the monodromy defect of $\sigma$: $\sigma \sim \sigma + 2\pi$, which requires the kink configuration and leads to the area law.
The situation of the spatial Wilson loop at the Higgs phase $(\sigma, \phi) = (\pi, 0)$ is parallel to that of the spatial 't Hooft loop at the (monopole-condensed) confining phase $(\sigma, \phi) = (0, \pi / 2)$.
Consequently, the screening mechanism via twist vortices, detailed in the main text, is crucial again for explaining the perimeter law of the spatial Wilson loop in the Higgs phase.
Thus, the operator detecting the spontaneous breakdown of $\left( \mathbb{Z}_N^{[1]} \right)_{\mathrm{3d}}$ will be $W(C) T_{\mathrm{twist}}(C)$ in the 3d description.

It would be important to observe how the Wilson-’t Hooft classification works for all possible gapped phases in $\mathcal{N}=1^*$ $SU(N)$ SYM on $\mathbb{R}^3\times S^1$.

    \item Appendix \ref{app:Lattice} offers an intuitive understanding of the screening mechanism realized by a twist vortex in a lattice model. 
    To explicitly address topological aspects within a lattice gauge theory, a more rigorous treatment, specifically utilizing the modified Villain lattice, would be necessary. 
    The development of a modified Villain lattice formulation for the $U(1)^{N-1} \rtimes S_N$ gauge theory is a promising direction for future research, generalizing Ref.~\cite{Jacobson:2024muj}. 

    \item One implication of this study is that twist vortex operators can play a crucial role when considering screening, particularly if the IR effective theory is an $S_N$-gauged (or $S_N$-quotiented) theory. For instance, when an $SU(N)$ gauge theory is compactified on a periodic $T^2$, it naively reduces to a sigma model whose target manifold is the moduli space of flat connections on $T^2$, given by $T^{2(N-1)}/S_N$ \cite{Yamazaki:2017ulc}. We expect that twist vortices may also become significant in such dimensional reduction scenarios.

    \item We have concentrated on the $\mathbb{R}^3\times S^1$ semiclassics in this paper.
    It would be interesting to investigate 't Hooft and dyonic loops in various other compactified setups, which consider the 't~Hooft twisted boundary condition on small $\mathbb{R}\times T^3$ \cite{vanBaal:2000zc, GarciaPerez:1992fj, Yamazaki:2017ulc, Cox:2021vsa, Poppitz:2022rxv, Gonzalez-Arroyo:2023kqv} and on small $\mathbb{R}^2\times T^2$ \cite{Tanizaki:2022ngt}. 
    For example, since one can smoothly interpolate between the monopole semiclassics on $\mathbb{R}^3\times S^1$ and the center-vortex semiclassics on $\mathbb{R}^2 \times T^2$ \cite{Hayashi:2024yjc, Hayashi:2024psa, Guvendik:2024umd}, the behavior of 't~Hooft loops in the latter case is readily predictable from the result of this paper.

    
\end{itemize}

\acknowledgments

The authors appreciate the YITP long-term workshop ``Hadrons and Hadron Interactions in QCD 2024'' (YITP-T-24-02) for providing the opportunities of useful discussions.  
This work was partially supported by Japan Society for the Promotion of Science (JSPS)
Research Fellowship for Young Scientists Grant No. 23KJ1161 (Y.H.), by JSPS KAKENHI
Grant No. 23K22489 (Y.T.), and by Center for Gravitational Physics and Quantum Information (CGPQI) at Yukawa Institute for Theoretical Physics.

\appendix

\section{'t Hooft loop in simplified lattice model on \texorpdfstring{$\mathbb{R}^3 \times S^1$}{R3xS1} }
\label{app:Lattice}

In this Appendix, we consider the 't Hooft loop in the $SU(N)$ lattice gauge theory on $\mathbb{R}^3 \times S^1$.
Our aim here is to present how the 't Hooft loop in the UV theory can be rewritten in the abelianized IR theory.
This simplified lattice model gives a clear understanding of the notion of screening by the twist vortex.

In the lattice gauge theory, the $\mathbb{Z}_N^{[1]}$ defect can be realized as the twist of the plaquette term.
Let $\Sigma$ be a surface in the dual lattice.
We can introduce the $\mathbb{Z}_N^{[1]}$ defect on $\Sigma$ by the following replacement in the plaquette action \cite{Kapustin:2014gua}:\footnote{
As long as the background $\mathbb{Z}_N$ plaquette gauge field is flat modulo $N$, we can consistently impose the admissibility condition by L{\"u}scher~\cite{Luscher:1981zq}, which allows us to study the topology of the $SU(N)$ link variables. 
Such topological properties under the presence of the flat background gauge fields are studied in \cite{Abe:2023ncy, Abe:2024fpt, Morikawa:2025ldq}. 
}
\begin{align}
    \operatorname{tr}(U_\Box) \to \operatorname{tr}(\rme^{-\frac{2 \pi \im}{N}}U_\Box), ~~\text{for} ~\Box\in \Sigma^*, 
\end{align}
where $\Sigma^*$ is a set of (original-lattice) plaquettes intersecting $\Sigma$.
With this definition, we can naturally define the 't Hooft loop, which is the $\mathbb{Z}_N^{[1]}$ defect on an open surface.
This is indeed the definition employed in \cite{deForcrand:2000fi}.

\subsection{Simplified lattice setup}

We consider these (spatial and temporal) 't Hooft loops in the following simplified $\mathbb{R}^3 \times S^1$ setup:
\begin{itemize}
    \item For simplicity, we assume that the temporal direction ($S^1$ direction) has only one link $N_4=1$, where the Polyakov loop becomes $P=U_4 \in SU(N)$.
    The link variable in the spatial directions is denoted by $U_\ell \in SU(N)$.

    This model can be described as the 3d $SU(N)$ gauge theory with the adjoint scalar $U_4$.
    The plaquette term in the $(\mu 4)$ plane $(\mu=1,2,3)$ can be translated as the hopping term of the adjoint scalar $U_4$.

    \item  Instead of adjoint fermions, we simply add a center-stabilizing potential $V_{\mathrm{eff}}(U_4)$ as the deformed Yang-Mills theory \cite{Unsal:2008ch}, which is minimized at the center-symmetric point. 
\end{itemize}

The action reads,
\begin{align}
    S[U_\ell,U_4] 
    &= -
    \frac{\beta}{2} \sum_{\Box:~ \mathrm{spatial}} \operatorname{tr}(U_\Box)
    - \frac{\beta}{2} \sum_{x, \mu=1,2,3}\, \operatorname{tr} \left[U_4(x) U_\mu(x) U_4^\dagger (x+\hat{e}_\mu) U_\mu^\dagger (x)\right] \notag \\
    &~~~~~~~~~~~~ + \sum_{n=1}^{[N/2]}\sum_xa_n |\operatorname{tr}U_4^n(x)|^2 + c.c.
\end{align}
where $\hat{e}_\mu$ denotes the unit vector in the spatial $\mu$ direction.
The deformation parameters $\{ a_n \}$ are introduced to stabilize the center symmetry.

To derive the low-energy effective theory as in the main text, we adopt the Polyakov gauge, which diagonalizes $U_4$, 
\begin{align}
    U_4 = \operatorname{diag}(\rme^{\im \varphi_1},\cdots,\rme^{\im \varphi_N}),
\end{align}
with the traceless constraints $\varphi_1 + \cdots + \varphi_N = 0~(\operatorname{mod}2\pi)$.
Then, at a generic point of $U_4$, the off-diagonal components of $U_\ell$ become massive.
Then, with the center-stabilizing deformation, the link variable $U_\ell$ will be abelianized: $SU(N) \rightarrow U(1)^{N-1}$.

More precisely, for the diagonal configuration $U_4 \in U(1)^{N-1}$, the hopping term of $U_4$ does not suppress the link variable $U_\ell$ that gives the automorphism of $U(1)^{N-1}$ (which is the normalizer of $U(1)^{N-1}$ in $SU(N)$).
Thus, we should keep $U(1)^{N-1} \rtimes S_N$ gauge for the link variable $U_\ell$.
As a low-energy effective theory, the discrete $S_N$ gauge should be flat, and can be fixed by restricting the holonomy into the Weyl chamber.
However, it will be convenient to keep the $S_N$ gauge for discussing the 't Hooft loop.

To parameterize $U(1)^{N-1} \rtimes S_N$ gauge, we write
\begin{align}
    U_\ell = P_{\sigma_\ell} \operatorname{diag}(\rme^{\im (a_\ell)_1},\cdots,\rme^{\im (a_\ell)_N}), \label{eq:SN_cartan_decomposition_lat}
\end{align}
where $P_{\sigma_\ell}$ is the $(N \times N)$ permutation matrix in $SU(N)$ representing $\sigma_\ell \in S_N$, and $(a_\ell)_j$ denotes the $U(1)^{N-1}$ factor with the constraint $(a_\ell)_1 + \cdots + (a_\ell)_N = 0~(\operatorname{mod}2\pi)$.

In this Appendix, we simply assume that the off-diagonal massive modes can be ignored, and consider the 3d $U(1)^{N-1} \rtimes S_N$ gauge theory with the $U(1)^{N-1}$-valued adjoint scalar (holonomy).
The deformation potential has minima at $U_4 = C$ and its permuted points; we mostly assume that the deformation potential is strong enough and constrains $U_4$ into the minima.
Then, we may regard $U_4$ as the $S_N$-valued scalar, that is, an $S_N$ higgs field.


\subsection{Spatial 't Hooft loop}

In this subsection, we address the following points:
\begin{itemize}
    \item \textbf{Motivation for the definition in the continuum}

    As discussed above, lattice gauge theory provides a natural definition of the 't Hooft loop. Based on this, we motivate the definition of the 't Hooft loop adopted in the continuum formulation in the main text, (\ref{eq:spatial_tHt_def}).

    \item \textbf{Perimeter law of the 't Hooft loop on the $U(1)^{N-1} \rtimes S_N$ lattice}

    We explain the perimeter law behavior of the 't Hooft loop in the $U(1)^{N-1} \rtimes S_N$ lattice gauge theory. A configuration with non-flat $S_N$ gauge field plays a crucial role in this mechanism.

    \item \textbf{Twist vortex and screened 't Hooft loop}

    As the $S_N$ gauge field is flat in the low-energy effective theory, we shall interpret the above mechanism where the $S_N$ flatness is imposed.
    For this purpose, we introduce the twist vortex, and the 't Hooft loop screened by a twist vortex reproduces the relevant configuration that leads to the perimeter law.
\end{itemize}
Here, we provide an intuitive explanation of these points, setting aside lattice subtleties. Our primary aim is to motivate the continuum definition and to offer an illustrative example of screening by a twist vortex.

\begin{figure}[h]
\centering
\includegraphics[width = 0.6 \linewidth]{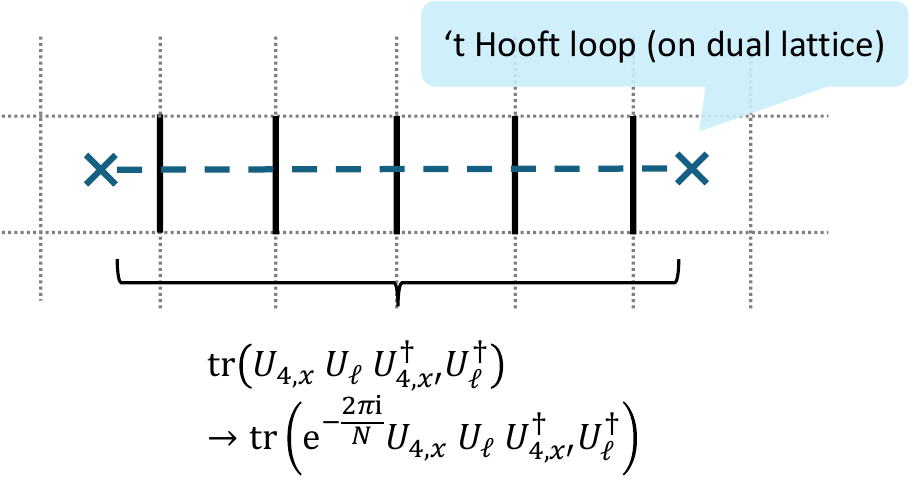}
\caption{
An illustration of the spatial 't Hooft loop $H(C;\Sigma)$ (in the 2d cross section) .
The $\mathbb{Z}_N$ twist is inserted in the hopping term which crosses the dual-lattice surface $\Sigma$.
}
\label{fig:lat_tHooftloop}
\end{figure}

The spatial 't Hooft loop $H(C;\Sigma)$ is constructed by the $\mathbb{Z}_N^{[1]}$ defect on an open surface $\Sigma$ on $\mathbb{R}^3$.
Note that the open surface $\Sigma$ is defined on the dual lattice.
In terms of the 3d original lattice (after dimensional reduction along the $S^1$ direction), the dual surface $\Sigma$ corresponds to a set of links:
\begin{align}
(\Sigma^*) =\{ \ell :\text{spatial link}~|~\text{the link }\ell \text{ intersects the surface }\Sigma\}.
\end{align}
Inserting the 't Hooft loop $H(C;\Sigma)$ is equivalent to the replacement of the hopping term:
\begin{align}
    \operatorname{tr} \left[U_4(x) U_\mu(x) U_4^\dagger (x+\hat{e}_\mu) U_\mu^\dagger (x)\right] \rightarrow \rme^{-\frac{2 \pi \im}{N}}\operatorname{tr}& \left[U_4(x) U_\mu(x) U_4^\dagger (x+\hat{e}_\mu) U_\mu^\dagger (x)\right] \notag \\
    &~~~\mathrm{for}~~(x,x+\hat{e}_\mu)\in \Sigma^*, \label{eq:lattice_spatial_tHooft}
\end{align}
where $(x,x+\hat{e}_\mu)$ denotes the link from $x$ to $x+\hat{e}_\mu$.
An illustration of the 't Hooft loop $H(C;\Sigma)$ (in the 2d cross section) is shown in Figure \ref{fig:lat_tHooftloop}.

In the abelianized description, this replacement can be written as,
\begin{align}
    \cos((d \varphi_j)_\ell) \to \cos((d \varphi_j)_\ell- 2\pi/N).
\end{align}
This defect is nothing but the $2\pi/N$-shift defect of the holonomy on the open surface, which is a straightforward definition of the 't Hooft loop.

In terms of the continuum language, this defect represents a superposition of the shift defect of the holonomy $\vec{\phi} \mapsto \vec{\phi} - 2 \pi \vec{\mu}_1 + 2 \pi \vec{\alpha}$ for all $\vec{\alpha} \in \Lambda_{\mathrm{roots}}$.
Note that this superposition is inevitable because the Wilson lattice does not control the winding number.
To find a continuum counterpart, we may extract the dominant $N$ terms, which are the shift defects of $\vec{\phi} \mapsto \vec{\phi} - 2 \pi \vec{\nu}_j$ for $j=1,\cdots,N$, where $\vec{\nu}_j$ denotes the weight vector of the fundamental representation.
This indeed represents the 't Hooft loop $H_{IR}(C;\Sigma)$ introduced in (\ref{eq:spatial_tHt_def}).

Naively, as the holonomy is subject to the deformation potential, this 't Hooft loop would exhibit the area-law falloff, even in the confining phase. 
In the IR effective theory (where the flatness condition of $S_N$ gauge is assumed), this defect requires a kink of the holonomy spanning the surface.

As described in the main text, screening effects can determine whether an extended object follows a perimeter law or an area law. 
When heavy degrees of freedom are integrated out, the extended objects can be dressed by line operators that are dynamical in the UV theory. 
Although such dressing terms typically do not contribute to the low-energy effective theory in the bulk, their contribution, while suppressed by the UV mass scale, scales according to a perimeter law. 
Consequently, this dressing affects the determination of whether the extended objects obey an area law or a perimeter law.

The key mechanism is the screening by the twist vortex.
After this screening, we can write the 't Hooft loop within the low-energy effective theory, i.e., in terms of the 3d $U(1)^{N-1}$ gauge theory.

\subsubsection{Perimeter law in the \texorpdfstring{$U(1)^{N-1} \rtimes S_N$}{U1**(N-1)xSN} Wilson lattice}

Before discussing the screening by the twist vortex, let us consider the spatial 't Hooft loop in the $U(1)^{N-1} \rtimes S_N$ Wilson lattice.
In this formulation, the flatness condition is not strictly imposed, which implies that the twist vortex is heavy but dynamical.
We will demonstrate that the spatial 't Hooft loop exhibits the perimeter law in this setup.

The deformation potential favors $U_4 = C$ and its permuted configurations.
Suppose that we take $U_4 = C$ on one side of the surface $\Sigma$ by choosing a gauge.
Then, if the link variable $U_\ell$ across $\Sigma$ has no $S_N$ component $P_{\sigma_\ell} =1$, then $U_4$ should be $\rme^{-2\pi \im /N} C$ on the other side of the surface $\Sigma$.
We can compensate for this change by the permutation, as $\rme^{-2\pi \im /N} C = S^{-1} C S$.
Hence, we can keep $U_4 = C$ if we choose the $S_N$ component of the link variable as $P_{\sigma_\ell} = S^{-1}$ for $\ell \in (\Sigma)^*$.

If the flatness condition for the discrete part is imposed, this configuration is not consistent.
It is impossible to end the nontrivial discrete gauge $P_{\sigma_\ell} = S^{-1}$ on the boundary $C$ within the low-energy description.

In the $U(1)^{N-1} \rtimes S_N$ Wilson lattice, the endpoint of the $S_N$ gauge transformation is allowed.
Even though this endpoint yields a heavy action cost, this action cost only appears on the endpoint, which is the boundary $C$. 
Thus, this action cost only gives the perimeter-law falloff.

\begin{figure}[h]
\centering
\includegraphics[width = 0.75 \linewidth]{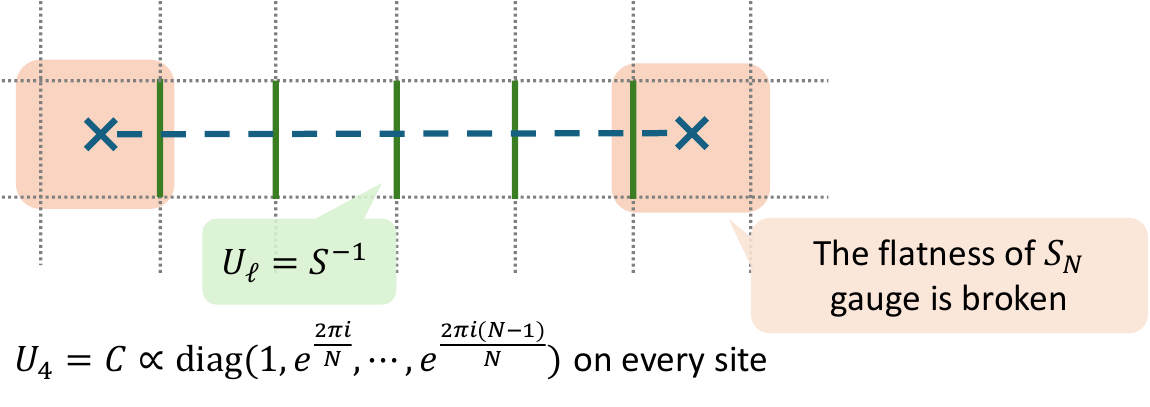}
\caption{
A configuration which gives the perimeter law of the spatial 't Hooft loop $H(C;\Sigma)$ (in the 2d cross section).
If we take $U_4 = C$, the $S_N$ gauge configuration $U_\ell = S^{-1}$ for $\ell \in (\Sigma)^*$ can minimize both the $U_4$-hopping term and the deformation potential.
The flatness condition of the spatial plaquette is broken on the boundary $C$, which only gives the perimeter law.
}
\label{fig:lat_SNgauge}
\end{figure}

In summary, it is possible to fix $U_4 = C$ by using the $S_N$ gauge configuration (which breaks the flatness on the boundary $C$). This configuration yields the perimeter-law falloff of the 't Hooft loop.
A cartoon of this configuration is shown in Figure \ref{fig:lat_SNgauge}.

This is a realization of the screening by the twist vortex.
Since the flatness condition of the discrete gauge is imposed in the low-energy effective theory, this configuration is not included in the low-energy theory.
To implement this configuration within the low-energy effective theory, we have to invoke the notion of screening, as discussed in Section \ref{sec:Screening_tHooftloop}.
In what follows, we construct the twist vortex, and we show that the screened 't Hooft loop represents the above configuration (Figure \ref{fig:lat_SNgauge}) in the setup where the $S_N$ flatness is imposed.

\subsubsection{Twist vortex}

Now, we will see the lattice counterpart of the ``screening by the twist vortex.''
Let us construct an operator in the 3d $U(1)^{N-1} \rtimes S_N$ lattice gauge theory, which corresponds to the twist vortex in the low-energy effective theory\footnote{For a well-controlled lattice setup, we should employ the Villain lattice.
Note that the twist vortex in the $U(1) \rtimes S_2$ modified Villain lattice is extensively investigated in Ref.~\cite{Jacobson:2024muj}.
Here, we do not enter into the details of the explicit lattice construction of the twist vortex.
We work in the Wilson lattice, and just assume the existence of the low-energy effective theory.
A detailed analysis of the $U(1)^{N-1} \rtimes S_N$ modified Villain lattice would merit further study.
}.

The twist vortex corresponds to the boundary of an $S_N$ gauge transformation.
As the Gukov-Witten defect in the Wilson lattice, we can define the twist vortex as the twist of the spatial plaquette term.
Due to the gauge invariance (within $U(1)^{N-1} \rtimes S_N$ gauge), we will introduce the twist in terms of the conjugacy class of $U(1)^{N-1} \rtimes S_N$.
We focus on the conjugacy class of the cyclic Weyl permutation, $[P_W] = \{ \tau P_W \tau^{-1}~|~ \tau \in S_N \}$.

Here, we adopt the following definition.
The twist vortex is located on a line $C'$ in the 3d dual lattice, which corresponds to a set of spatial plaquettes. For such plaquettes $\Box\in (C')^*$, we replace the Boltzmann factor with
\begin{align}
    \rme^{-\frac{\beta}{2} \operatorname{tr}(U_\Box) + c.c.}
     &\to \int_{U(1)^{N-1} \rtimes S_N} dh~\rme^{-\frac{\beta}{2} \operatorname{tr}((h^{-1} S h)U_\Box) + c.c.} \label{eq:def_lat_twisted_vortex} 
\end{align}
In the low-energy theory, this defect requires that the $S_N$ component of the plaquette $U_\Box$ should be one permutation of the conjugacy class $[P_W]$.
In other words, when the (flat) $S_N$ discrete gauge theory is viewed as a defect network, a defect of a certain class can terminate at the twist vortex.


\subsubsection{Screening by twist vortex}

Now, let us consider the 't Hooft loop screened by the twist vortex $H(C;\Sigma)T_{\mathrm{twist}}(C)$.

Note again that, although the twist vortex is not a dynamical object in the IR effective theory, it is dynamical in the UV theory.
In the UV theory, the flatness condition for the $S_N$ gauge is not imposed as we have seen above, and such configurations represent the insertions of the twist vortex in terms of the IR effective theory.

In the low-energy effective theory, the $S_N$ gauge should be flat except for the twist vortex. 
By the $S_N$ gauge fixing, we can eliminate all unnecessary $S_N$ part of the link variables.
In the presence of the twist vortex, the only necessary $S_N$ gauge field is a cyclic Weyl permutation defect that spans an open surface $\Sigma'$ satisfying $\partial \Sigma' = C$.
The open surface can be deformed by gauge transformations, so we can choose $\Sigma' = \Sigma$.

The twist vortex induces an ensemble of $S_N$ gauge configurations labeled by an element of the conjugacy class including $P_W$, $[P_W]$:
\begin{align}
    \Biggl\{
    \sigma_\ell = \begin{cases}
        \tau &\text{ for }\ell \in  (\Sigma)^*\\
        1  &\text{ for }\ell \notin  (\Sigma)^*
    \end{cases}
    ~~~\Bigg|~\tau \in [P_W]
    \Biggr\}
    \label{eq:twisted_vortex_induced_SN_gauge}
\end{align}


To sum up, in the low-energy effective theory, the composite defect $H(C;\Sigma)T_{\mathrm{twist}}(C)$ induces
\begin{itemize}
    \item $H(C;\Sigma)$: twist in the hopping term of $U_4$ on the links in $(\Sigma)^*$, (\ref{eq:lattice_spatial_tHooft})
    \item $T_{\mathrm{twist}}(C)$: ensemble of the $S_N$ gauge configurations (\ref{eq:twisted_vortex_induced_SN_gauge}) with equal weights.
\end{itemize}
Now, for simplicity, let us pick a configuration $U_4 = C$, which is a minimum of the deformation potential.
The choice of the minimum can be arbitrary because we keep the explicit permutation invariance.
In particular, the ensemble generated by $T_{\mathrm{twist}}(C)$ possesses the following $S_N$ gauge configuration
\begin{align}
    P_{\sigma_\ell} = \begin{cases}
        S^{-1} &\text{ for }\ell \in  (\Sigma)^*\\
        1  &\text{ for }\ell \notin  (\Sigma)^*
    \end{cases}
\end{align}
as $P_W^{-1} \in [P_W]$.
In this configuration, the holonomy can be fixed at $U_4 = C$, and no extra action cost is required.
This configuration corresponds precisely to the case exhibiting the perimeter law (Figure \ref{fig:lat_SNgauge}).
Hence, we find that $\braket{H(C;\Sigma )T_{\mathrm{twist}}(C)}_{\mathrm{IR}}$ behaves as the perimeter law in the low-energy effective theory where the $S_N$ flatness is imposed.

When we fix $U_4 = C$ and write the 3d effective $U(1)^{N-1}$ gauge theory, the screened 't Hooft loop $H(C;\Sigma )T_{\mathrm{twist}}(C)$ simply becomes the cyclic Weyl permutation defect on the open surface $\Sigma$ for the spatial gauge field.
This is the lattice counterpart of Section \ref{sec:Screening_tHooftloop}.

\subsection{Temporal 't Hooft loop}
\label{app:lat_temp_tHloop}

We can also consider the temporal 't Hooft loop that extends in the $S^1$ direction.
The loop becomes a point, and the topological surface becomes a line in the 3d theory.
In the 3d lattice model, the non-genuine temporal 't Hooft loop $H(\{p,p'\};\gamma)$ corresponds to the following defect defined on the dual-lattice open line $\gamma$:
\begin{align}
    \operatorname{tr}(U_\Box) \to \operatorname{tr}(\rme^{-\frac{2 \pi \im}{N}}U_\Box), ~~\text{for} ~\Box\in \gamma^*, \label{eq:temp_tH_Wilsonlattice}
\end{align}
where $\gamma^*$ is a set of (original-lattice) plaquettes that intersect with the dual-lattice line $\gamma$.

Since this definition does not involve $U_4$, we can restrict ourselves to $U_4 = C$ as usual, and the setup can be reduced to the 3d $U(1)^{N-1}$ gauge theory.
Whether the temporal 't Hooft loop defined here obeys a perimeter law or an area law depends on the details of the dynamics. 
In this subsection, instead of this, we aim to motivate the definition of the temporal 't Hooft loop in the main text (\ref{eq:def_temp_tHloop}) from the perspective of the above lattice definition.

The continuum definition in the main text is written in terms of the dual photon. Thus, in this section, we intuitively translate the Wilson lattice into a Villain-like notation, where the field strength $\vec{f}$ is defined on plaquettes and the dual photon $\vec{\sigma}$ on dual lattice sites, and discuss how the temporal 't Hooft loop can be described in terms of the dual photon. 
Note that we only give an intuitive explanation for the continuum definition (\ref{eq:def_temp_tHloop}), ignoring lattice details\footnote{To treat the dual photon on the lattice exactly, we should work in the modified Villain lattice formalism, which will be out of the scope of this paper.}.


From the Wilson-lattice definition (\ref{eq:temp_tH_Wilsonlattice}), for plaquettes intersecting with $\gamma$ ($\Box \in \gamma^*$), the twisted plaquette term favors $U_{\Box} \approx \rme^{\frac{2 \pi \im}{N}}$.
In terms of the field strength $\vec{f}_{\Box}$ within the abelianized description, the $\mathbb{Z}_N^{[1]}$ defect will correspond to introducing the background:
\begin{align}
    \vec{f}_{\Box} \mapsto \vec{f}_{\Box} + 2 \pi \vec{\nu}_j
\end{align}
with any $j=1,\cdots,N$.
In principle, $U_{\Box} \approx \rme^{\frac{2 \pi \im}{N}}$ can mix with higher magnetic flux $2 \pi \vec{\nu}_j + 2\pi \vec{\alpha}~~(\vec{\alpha} \in \Lambda_{\mathrm{roots}})$, but we only extract minimal terms here.

Let us remember that the dual photon is an auxiliary field that imposes the Bianchi identity:
\begin{align}
    \rme^{\frac{\im}{2\pi} \sum_{\ell: \text{dual link} } (d\vec{\sigma})_{\ell} \vec{f}_{\star \ell}},
\end{align}
where $\star \ell$ is a plaquette intersecting with the dual-lattice link $\ell$.
We can guess that the non-genuine temporal 't Hooft loop $H(\{p,p'\};\gamma)$ would be
\begin{align}
    ``\left(   \sum_{j=1}^N \rme^{\im \int_\gamma \vec{\nu}_j \cdot  \sum_{\ell \in \gamma} (d\vec{\sigma})_{\ell}}   \right)",
\end{align}
However, this expression is naive; careful consideration of the endpoints is required.

Whereas the monopole is not well-controlled in the Wilson lattice, the Bianchi identity exactly holds in the Villain-type formulation.
To handle this difference and to remove the Bianchi-identity constraint, we attach
\begin{align}
    \sum_{\vec{\alpha} \in \Lambda_{\mathrm{roots}}} \rme^{\im \vec{\alpha}\cdot \vec{\sigma}_{s}},
\end{align}
for each dual-lattice site $s$ (or original-lattice cube).
Usually, the monopole requires a large action cost at the weak-coupling (will be singular in the continuum), and the Bianchi identity approximately holds.
At the endpoints of $\gamma$, the Bianchi identity would be easily broken, so we should treat this point more carefully.

Let us focus on one term with magnetic flux $2 \pi \vec{\mu}_1$, for simplicity.
We take a point $p \in \partial \gamma$ with the outgoing background magnetic flux $2 \pi \vec{\mu}_1$.

We first consider the trivial sector, where the Bianchi identity $(d\vec{f})_{\mathrm{cube}} = 0$ is imposed.
Then, due to the conservation of the magnetic flux, we need the incoming flux $2 \pi \vec{\mu}_1$ from the dynamical field, which leads to an extra action cost.
Even in the trivial sector, the presence of the background effectively forces the dynamical magnetic flux to exhibit a nontrivial divergence.
Thus, other sectors with nontrivial divergences $(d\vec{f})_{\mathrm{cube}} = 2 \pi \vec{\alpha}$ ($\vec{\alpha} \in \Lambda_{\mathrm{roots}}$) may be comparable to the trivial sector.

The extra action cost is exactly the same as that of $(d\vec{f})_{\mathrm{cube}} = 2 \pi \vec{\alpha}_{j,1} = 2 \pi (\vec{e}_j - \vec{e}_1)$ for $j=2,\cdots,N$. 
Indeed, in the sector of $(d\vec{f})_{\mathrm{cube}} = 2 \pi \vec{\alpha}_{j,1}$, we need additional incoming flux $2 \pi \vec{\nu}_j$ from the fluctuation, in order to compensate for the outgoing background magnetic flux $2 \pi \vec{\mu}_1$.
From the permutation symmetry, the extra action costs are identical irrespective of $j = 1,2,\cdots,N$.
Hence, we should sum over these sectors in equal weights.

The above observation suggests that we should sum over a certain class of monopole operators $\rme^{\im \vec{\alpha}\cdot \vec{\sigma}}$ at the endpoints:
\begin{align}
      \sum_{k=1}^N \left[ \left( \sum_{j=1}^{N} \rme^{-\im  \vec{\alpha}_{j,k} \cdot \vec{\sigma}(p)} \right) \rme^{\im \int_\gamma \vec{\nu}_k \cdot d \vec{\sigma}} \left( \sum_{j'=1}^{N} \rme^{\im \vec{\alpha}_{j',k} \cdot \vec{\sigma}(p')} \right)  \right],
\end{align}
where the open line $\gamma$ is from $p$ to $p'$.
In this expression, the first sum over the permutations is redundant, and we can reproduce the continuum definition:
\begin{align}
     H(\{p,p'\};\gamma) = \sum_{j,j'=1}^{N}  \rme^{\im  \vec{\alpha}_{j',j} \cdot \vec{\sigma}(p)} \rme^{\im \int_\gamma \vec{\nu}_j \cdot d \vec{\sigma}}  , \tag{\ref{eq:def_temp_tHloop}}
\end{align}
with $\vec{\alpha}_{j',j} = \vec{e}_{j'} - \vec{e}_{j}$.
Intuitively, this expression takes the form of
\begin{align}
       `` \left[ \left( \sum_{j=1}^{N} \rme^{-\im  \vec{\nu}_{j} \cdot \vec{\sigma}(p)}\right) \times \left(\mathbb{Z}_N\text{ topological line on } \gamma \right) \times \left( \sum_{j'=1}^{N} \rme^{\im  \vec{\nu}_{j} \cdot \vec{\sigma}(p')}\right) \right]",
\end{align}
which is not an appropriate expression since $\vec{\sigma}$ has the weight-vector periodicity.

This form is indeed what we intuitively expect as the temporal 't Hooft loop in the 3d monopole semiclassics.
In the 3d monopole semiclassics, the monopole operator is the $\rme^{\im \vec{\alpha}\cdot \vec{\sigma}}$ with a root vector $\vec{\alpha}$, so the non-genuine 't Hooft loop will be the weight-vector charge monopole operator attached to the topological line.

Let us summarize our findings in this section.
We have considered the simplified lattice model: 4d $SU(N)$ lattice gauge theory on $\{\text{3d lattice} \} \times \{\text{one link } (S^1)\}$ with the large deformation potential which restricts $U_4$ to the center symmetric points. By taking the Polyakov gauge, we can regard $U_4$ as an $S_N$-valued scalar: $U_4 = C$ and its permuted ones.
Ignoring the off-diagonal massive modes, we can write the 3d Wilson-type $U(1)^{N-1} \rtimes S_N$ gauge theory with the $S_N$-valued scalar $U_4$.

\begin{itemize}
    \item First, we consider the spatial 't Hooft loop in the 3d $U(1)^{N-1} \rtimes S_N$ Wilson lattice.
    Unlike the continuum case, the spatial 't Hooft loop shows the perimeter law (See Figure \ref{fig:lat_SNgauge}).

    The non-flat configuration of $S_N$ plays an important role in this argument.
    This is precisely the screening by the dynamical twist vortex.

    \item We then construct the operator corresponding to the twist vortex.
    The composite operator $H(C;\Sigma)T_{\mathrm{twist}}(C)$ possesses a term that acts nontrivially in the low-energy effective theory, where the flatness of $S_N$ gauge is imposed.
    In terms of the dual photon, the screened operator $H(C;\Sigma)T_{\mathrm{twist}}(C)$ is the cyclic Weyl permutation defect on the open surface $S$.

    \item We also consider the temporal 't Hooft loop, and present an intuitive argument to motivate the continuum definition (\ref{eq:def_temp_tHloop}) from observing the lattice definition (\ref{eq:temp_tH_Wilsonlattice}).
\end{itemize}

\bibliographystyle{utphys}
\bibliography{./QFT,./refs}

\end{document}